\documentclass[aps,prb,showpacs,twocolumn]{revtex4-1}
\usepackage{amssymb}
\usepackage{amsmath}
\usepackage{graphicx}
\usepackage{epsfig}

\begin{document}

\title{Dynamical bulk-edge correspondence for nodal lines in parameter space}
\author{R. Wang${}^1$, C. Li${}^1$, X. Z. Zhang${}^2$ and Z. Song${}^1$}
\email{songtc@nankai.edu.cn}
\affiliation{${}^1$School of Physics, Nankai University, Tianjin 300071, China \\
${}^2$College of Physics and Materials Science, Tianjin Normal University,
Tianjin 300387, China}

\begin{abstract}
Nodal line in parameter space, at which the energy gap closes up, can either
be the boundary separating two topological quantum phases or two
conventional phases. We study the topological feature of nodal line in
parameter space via a dimerized Kitaev spin chain with staggered transverse
field, which can be mapped onto the system of dimerized spinless fermions
with $p$-wave superconductivity. The quantum phase boundaries are straight
crossing degeneracy lines in 3D parameter space. We show that the nodal line
acts as a vortex filament associated with a vector field, which is generated
from the Zak phase of Bogoliubov-de Gennes band. We also investigate the
topological invariant in Majorana fermion representation for open chain. The
Majorana edge modes are not zero mode, but is still protected by energy gap.
The exact mid-gap states of the Majorana lattice allows to obtain the
corresponding Majorana probability distribution for whole 3D parameter
space, which can establish a scalar field to identify the topological
feature of the nodal lines. Furthermore, when we switch on a weak tunneling
between two ends of the Majorana lattice, the topological invariants of the
nodal lines can be obtained by the pumped Majorana probability from an
adiabatic passage encircling the lines. Numerical simulations of
quasi-adiabatic time evolution is performed in small system. We compute the
current to monitor the transport of Majorana fermion, which exhibits evident
character of topological pumping. Our results link the bulk topological
invariant to the dynamical behavior of Majorana edge mode, extending the
concept of bulk-edge correspondence.
\end{abstract}

\maketitle


\section{Introduction}

The quantum phase in condensed matter was believed to originate from the
symmetries atomic array and the phase boundary can be interpreted by Landau
symmetry breaking theory \cite{L.D1,L.D2} until the discovery of the quantum
Hall effect \cite{Thouless,von}. The topological features of matter have
attracted intensive studies in many aspects. From the spectral point of
view, there are gapped and gapless topological phases. The topological phase
with energy gap can be classified as two types \cite{X.Chen}, the one with
intrinsic topological order, which do not require any symmetry, such as
fractional quantum Hall systems \cite{R.B}, chiral spin liquids \cite{V.K}
and $Z_{2}$ spin liquids \cite{N.R}, and the one with symmetry protected
topological order \cite{F.P,X.Chen2,A.M,X-G,X.Chen3,O.M,Z-C}, such as the
spin-$1$ Haldane chain \cite{F.D.M} or $Z_{2}$ topological insulators \cite%
{C.L,C.L2,B.A,M.K}.

Besides the gapped topological phase, robust topological properties can also
exist for the systems that have stable Fermi points or nodal lines in the
energy spectrum. Examples of these gapless systems are Dirac points in
graphene \cite{K.N,M.F,A.H,Y.N,Y.K,Y.N2}, the A phase of superfluid 3He \cite%
{G.E,Y.T}, Weyl semimetals \cite%
{S.M,X.Wan,G.Xu,A.B,W.W,G.Chen,H-J,P.H,T.M,H.Weng,J.Liu}, and nodal
noncentrosymmetric superconductors \cite%
{F.Wang,M.S,B.B,A.P,P.M.R,A.P2,M.S2,K.Y,Y.T2,M.S3,S.M2,P-Y}.

The topology of such gapless state originates from the topological feature
of the nodal points and nodal lines, as vortex and vortex line in Brillouin
zone \cite%
{Wan,Yang,Burkov,Xu,Kim,Weng,Huang,Young,Wang,Wang1,Hou,Sama,Liu,Neupane,SXu,Lv,Lu,LC3,WP,W.Chen,Xiao-Qi}%
. However, it is a little hard to measure the topological invariant in
experiments, which is defined under the periodic boundary condition.
Fortunately, there is a particularly important concept, the bulk-edge
correspondence, which indicates that, a nontrivial topological invariant in
the bulk indicates that localized edge modes only appear in the presence of
opened boundaries in the thermodynamic limit \cite{C.L,B.A,Y.H,A.Y,S.Ryu,X-L}%
.

In this paper, we explore an alternative topological character hidden in the
condensed matter. We study the feature of nodal line in parameter space via
a concrete system. Such a nodal line as quantum phase boundary, is different
from the nodal line in the momentum space. We investigate a dimerized Kitaev
spin chain with a staggered transverse field, which can be mapped onto the
system of dimerized spinless fermions with $p$-wave superconductivity. The
quantum phase boundaries are straight crossing lines in the $3$D parameter
space. We establish a vector field \textbf{P} in a parameter space,
generated from the Zak phase of Bogoliubov-de Gennes band based on the exact
solution. We show that the phase boundary line acts as\ a vortex filament
associated with the field \textbf{P}, which reveals the topological feature
of the nodal line. By Majorana transformation, we also study the topological
pump corresponding to the \textbf{P} field. We investigate a Thouless
quantum pump with Majorana fermion. Analytical and numerical results show
that the quantized pumping charge is equivalent to the topological invariant
of the nodal line and can be obtained via a quasi-adiabatic process in a
weak open system, which reveals a dynamical bulk-edge correspondence.

This paper is organized as follows. In section \ref{Hamiltonians and phase
diagram}, we present the model and phase diagram. In section \ref{Band
structure and symmetry}, we investigate the band structure and symmetries.
Section \ref{Topological invariants} devotes to the topological invariant of
the nodal lines. In section \ref{Edge mode and adiabatic transport}, we
study the Majorana representation of the model.\ Section \ref{Summary}
summarizes the results and explores its implications.

\section{Hamiltonians and phase diagram}

\label{Hamiltonians and phase diagram}

We start our investigation by considering a dimerized Kitaev spin ring with
spatially modulated real and complex fields%
\begin{equation*}
H=\sum_{j=1}^{N}\left( \lambda \sigma _{2j-1}^{x}\sigma _{2j}^{x}+\sigma
_{2j}^{y}\sigma _{2j+1}^{y}\right) +\sum_{j=1}^{2N}g_{j}\sigma _{j}^{z},
\end{equation*}%
where the external field $g_{j}$ can be either real or complex. We take $%
g_{j}=g_{o}$\ ($g_{e}$)\ for odd (even) $j$. Here $\sigma _{j}^{\alpha }$ ($%
\alpha =x,$ $y,$ $z$) are the Pauli operators on site $j$, and satisfy the
periodic boundary condition $\sigma _{j}^{\alpha }\equiv \sigma
_{j+2N}^{\alpha }$.

The ground state properties of this model with zero external field have been
studied recently\cite{FXY}, while its non-Hermitian version without
dimerization is investigated in Refs. \cite{WXH,ZQB}. It has been shown that
a second order quantum phase transition (QPT) occurs at $\lambda =\pm 1$\
but zero external field.

In the following, we will diagonalize the Hermitian model. It can perform
the Jordan-Wigner transformation \cite{P.Jordan}%
\begin{eqnarray}
\sigma _{2j}^{x} &=&-\prod\limits_{l<2j}\left( 1-2c_{l}^{\dagger
}c_{l}\right) \left( c_{2j}^{\dagger }+c_{2j}\right) ,  \notag \\
\sigma _{2j}^{y} &=&-i\prod\limits_{l<2j}\left( 1-2c_{l}^{\dagger
}c_{l}\right) \left( c_{2j}^{\dagger }-c_{2j}\right) ,  \notag \\
\sigma _{2j}^{z} &=&1-2c_{2j}^{\dagger }c_{2j},
\end{eqnarray}%
to replace the Pauli operators by the fermionic operators $c_{j}$. The
obtained fermionic Hamiltonian reads

\begin{eqnarray}
H &=&\sum_{j=1}^{N}\{[\lambda (c_{2j-1}^{\dagger }c_{2j}^{\dagger
}+c_{2j-1}^{\dagger }c_{2j})  \notag \\
&&+(c_{2j+1}^{\dagger }c_{2j}^{\dag }+c_{2j}^{\dag }c_{2j+1})+\mathrm{H.c.}]
\notag \\
&&+g_{o}(1-2c_{2j-1}^{\dagger }c_{2j-1})+g_{e}(1-2c_{2j}^{\dagger }c_{2j})\},
\end{eqnarray}%
which describes a system of dimerized spinless fermions with $p$-wave
superconductivity is a variant of spinless $p$-wave superconductor wire\
\cite{Kitaev}. Here we neglect the difference between even and odd number
parity, which has no effect on the quantum phase diagram in the large $N$
limit.

It is a bipartite lattice, i.e., it has two sublattices $A$, $B$ such that
each site on lattice $A$ has its nearest neighbors on sublattice $B$, and
vice versa. We introduce the Fourier transformations in two sub-lattices%
\begin{equation}
c_{j}=\frac{1}{\sqrt{N}}\sum_{k}e^{ikl}\left\{
\begin{array}{cc}
\alpha _{k}, & j=2l \\
\beta _{k}, & j=2l-1%
\end{array}%
\right. ,
\end{equation}%
where $l=0,1,2,...,N$, $k=2m\pi /N$, $m=0,1,2,...,N-1$. Spinless fermionic
operators in $k$ space $\alpha _{k},$\ $\beta _{k}$ are%
\begin{equation}
\left\{
\begin{array}{cc}
\alpha _{k}=\frac{1}{\sqrt{N}}\sum\limits_{j}e^{-ikl}c_{j}, & j=2l \\
\beta _{k}=\frac{1}{\sqrt{N}}\sum\limits_{j}e^{-ikl}c_{j}, & j=2l-1%
\end{array}%
\right. .
\end{equation}%
This transformation block diagonalizes the Hamiltonian due to its
translational symmetry, i.e.,%
\begin{eqnarray}
H^{\prime } &=&\sum_{k\in (0,\pi ]}H_{k}=\sum_{k\in (-\pi ,0]}H_{k}=\frac{1}{%
2}\sum_{k\in (-\pi ,\pi ]}H_{k}  \notag \\
&=&\psi _{k}^{\dagger }h_{k}\psi _{k},
\end{eqnarray}%
satisfying $\left[ H_{k},H_{k^{\prime }}\right] =0$. Here\ $H^{\prime }$ is
rewritten in the Nambu representation with the basis%
\begin{equation}
\psi _{k}=\frac{1}{\sqrt{2}}\left(
\begin{array}{c}
-\beta _{-k}^{\dagger }+\beta _{k} \\
-\alpha _{-k}^{\dagger }+\alpha _{k} \\
\alpha _{-k}^{\dagger }+\alpha _{k} \\
-\beta _{-k}^{\dagger }-\beta _{k}%
\end{array}%
\right) .
\end{equation}%
And $h_{k}$ is a $4\times 4$\ matrix

\begin{equation}
h_{k}=\left(
\begin{array}{cccc}
0 & 0 & \gamma _{-k} & g_{o} \\
0 & 0 & -g_{e} & 0 \\
\gamma _{k} & -g_{e} & 0 & 0 \\
g_{o} & 0 & 0 & 0%
\end{array}%
\right) ,  \label{h_k}
\end{equation}%
where $\gamma _{k}=\lambda +e^{ik}$. The eigenstates of $h_{k}$ are $%
\left\vert \phi _{\rho \sigma }^{k}\right\rangle $\ $\left( \sigma ,\rho
=\pm \right) $ with eigenvalues%
\begin{equation}
\varepsilon _{\rho \sigma }^{k}=\frac{\rho }{\sqrt{2}}\sqrt{\Lambda
_{k}+\sigma \sqrt{\Lambda _{k}^{2}-4g_{o}^{2}g_{e}^{2}}},  \label{spectrum}
\end{equation}%
where $\Lambda _{k}=\left\vert \gamma _{k}\right\vert
^{2}+g_{o}^{2}+g_{e}^{2}$. The explicit expression of $\left\vert \phi
_{\rho \sigma }^{k}\right\rangle $\ is%
\begin{equation}
|\phi _{\rho \sigma }^{k}\rangle =\frac{1}{\Omega _{\rho \sigma }}\left(
\begin{array}{c}
\varepsilon _{\rho \sigma }^{k}g_{e}\gamma _{-k} \\
\varepsilon _{\rho \sigma }^{k}[\left( \varepsilon _{\rho \bar{\sigma}%
}^{k}\right) ^{2}-g_{e}^{2}] \\
\text{ }g_{e}[\left( \varepsilon _{\rho \sigma }^{k}\right) ^{2}-g_{o}^{2}]
\\
g_{e}g_{o}\gamma _{-k}%
\end{array}%
\right) ,  \label{vectors}
\end{equation}%
where $\Omega _{\rho \sigma }=\rho \sqrt{2}g_{e}(\varepsilon _{\rho \sigma
}^{k})^{-1}$ $\{[(\varepsilon _{\rho \sigma }^{k})^{4}-(g_{e}g_{o})^{2}]$ $%
[(\varepsilon _{\rho \sigma }^{k})^{2}-g_{o}^{2}]\}^{\frac{1}{2}}$ are
normalization factors.

There are four\ Bogoliubov-de Gennes\ bands from the eigenvalues of $h_{k}$,
indexed by $\rho ,\sigma =\pm $. The band touching points for $\sigma =\pm $%
\ occur at $k_{c}$\ when%
\begin{equation}
\Lambda _{k_{c}}^{2}-4g_{o}^{2}g_{e}^{2}=0,
\end{equation}%
or for $\rho =\pm $, when
\begin{equation}
\Lambda _{k_{c}}=\sqrt{\Lambda _{k_{c}}^{2}-4g_{o}^{2}g_{e}^{2}}.
\end{equation}%
The solutions of above equations form the degeneracy lines%
\begin{equation}
\lambda =\pm 1,g_{o}=\pm g_{e},
\end{equation}%
and planes indicates by%
\begin{equation}
g_{o}=0,
\end{equation}%
or%
\begin{equation}
g_{e}=0,
\end{equation}%
in parameter space $(g_{e},g_{o},\lambda )$, which are the quantum phase
boundaries. In this paper, we only focus on the nodal lines. We plot the
degeneracy lines in parameter space $(g_{e},g_{o},\lambda )$\ in Fig. \ref%
{fig1}. We see that there are four nodal lines in the $3$-dimensional
parameter space, which cross at the $\lambda \ $axis with$\ \lambda =\pm 1$.
In the planes of $g_{o}=\pm g_{e}$, four nodal lines are boundaries between
trivial and non-trivial topological phases \cite{Ning}. Whereas in other
surface containing a nodal line, there are no guarantee for the occurrence
of a topological QPT when crossing the nodal line.\textbf{\ }
\begin{figure}[tbp]
\includegraphics[ bb=40 327 536 777, width=0.4\textwidth, clip]{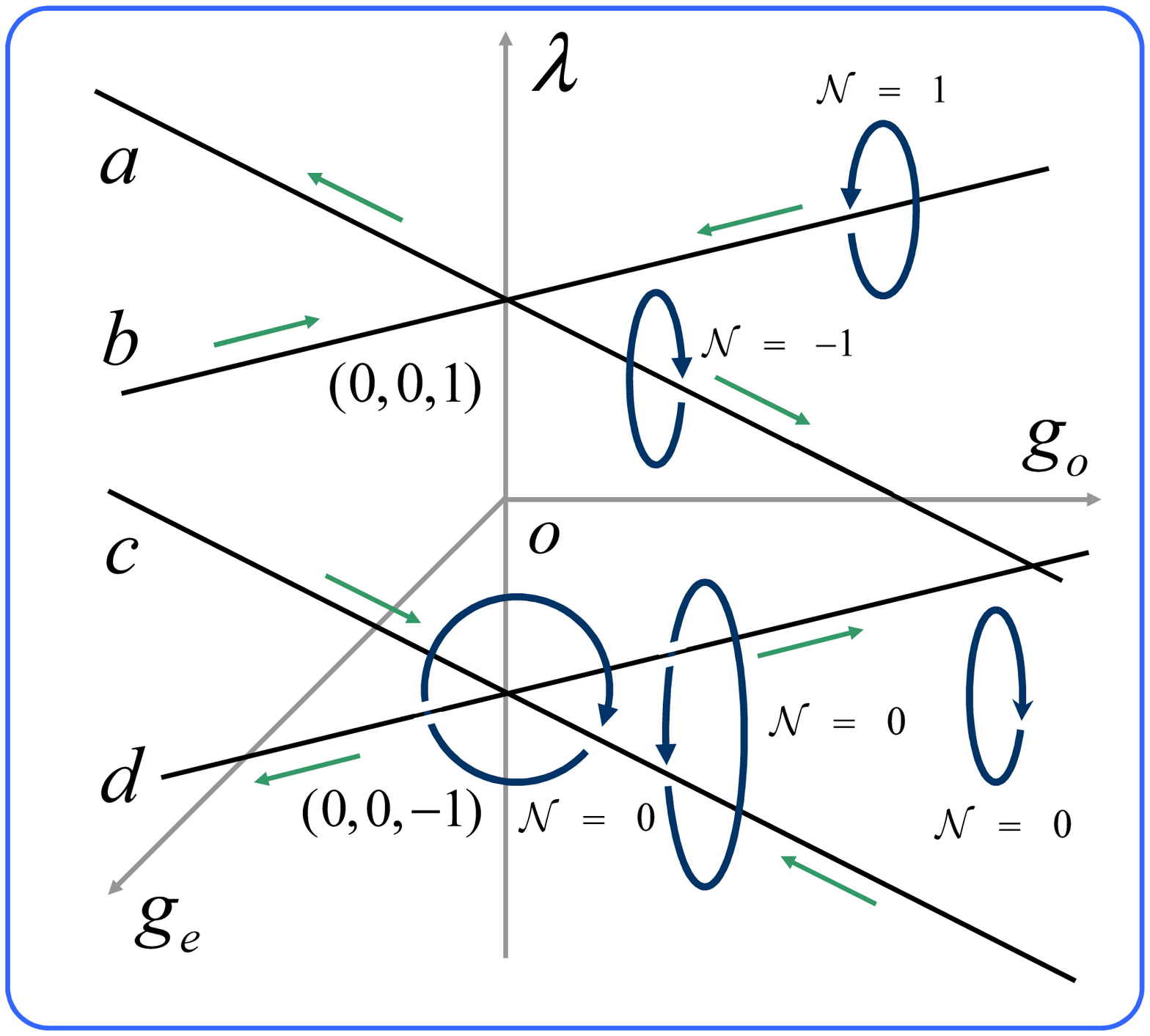}
\caption{Schematic illustration of the degenerate lines and corresponding
topological invariant in $3$-dimensional parameter space. There are four
lines a, b, c, d, which cross at the $\protect\lambda $ axis. The
topological feature of the nodal lines is characterized by the path integral
along the loops, as topological invariant, which is denoted for each loop.
The obtained results indicate that the topological feature of the degenerate
lines is equivalent to that of the magnetic fields produced by the
current-carrying wires (a,b,c,d) with directions denoted by arrows. }
\label{fig1}
\end{figure}

\begin{figure}[tbp]
\begin{minipage}{0.48\linewidth}
\centerline{\includegraphics[width=4.6cm]{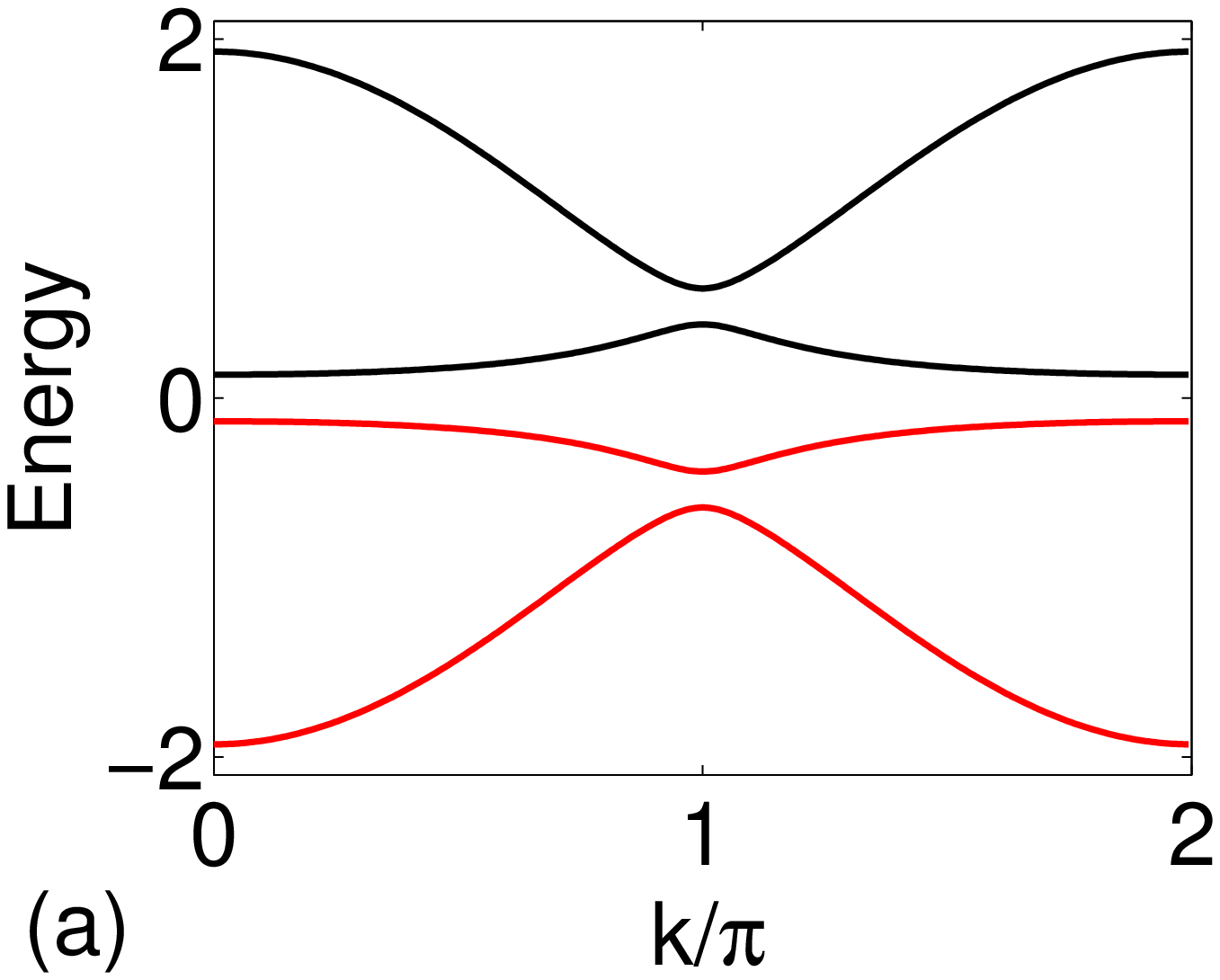}}
\end{minipage}
\begin{minipage}{0.48\linewidth}
\centerline{\includegraphics[width=4.6cm]{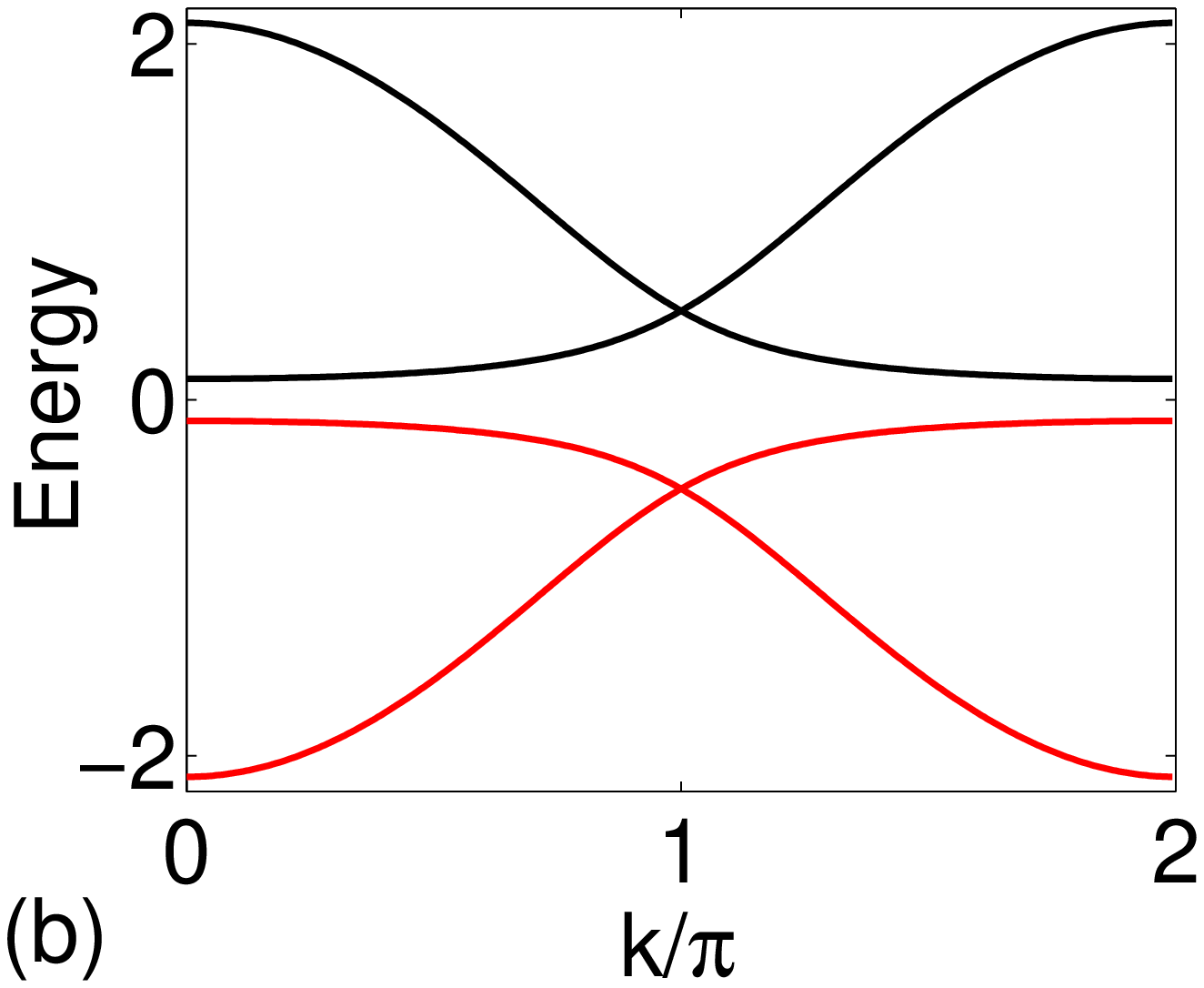}}
\end{minipage}
\begin{minipage}{0.48\linewidth}
\centerline{\includegraphics[width=4.6cm]{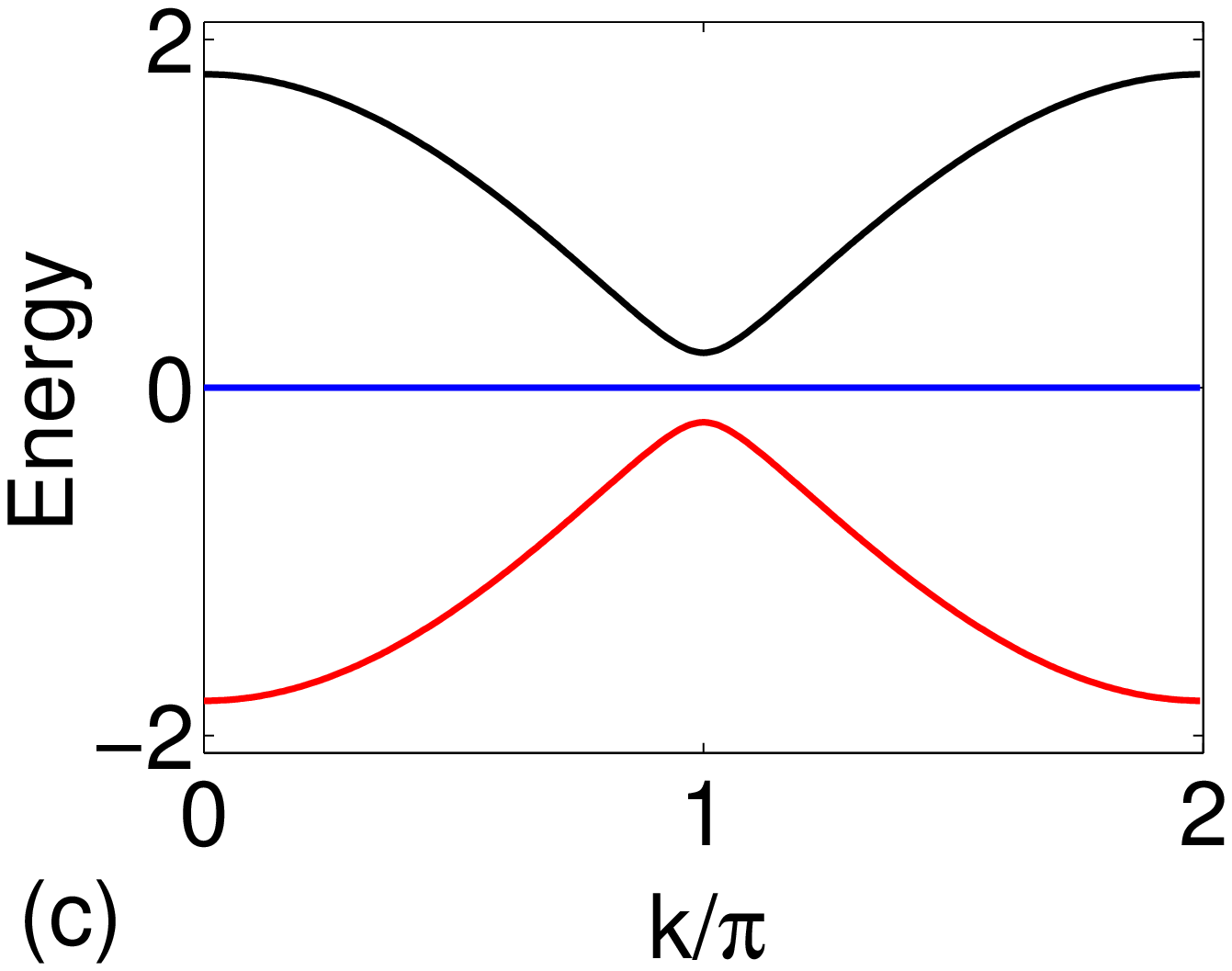}}
\end{minipage}
\begin{minipage}{0.48\linewidth}
\centerline{\includegraphics[width=4.6cm]{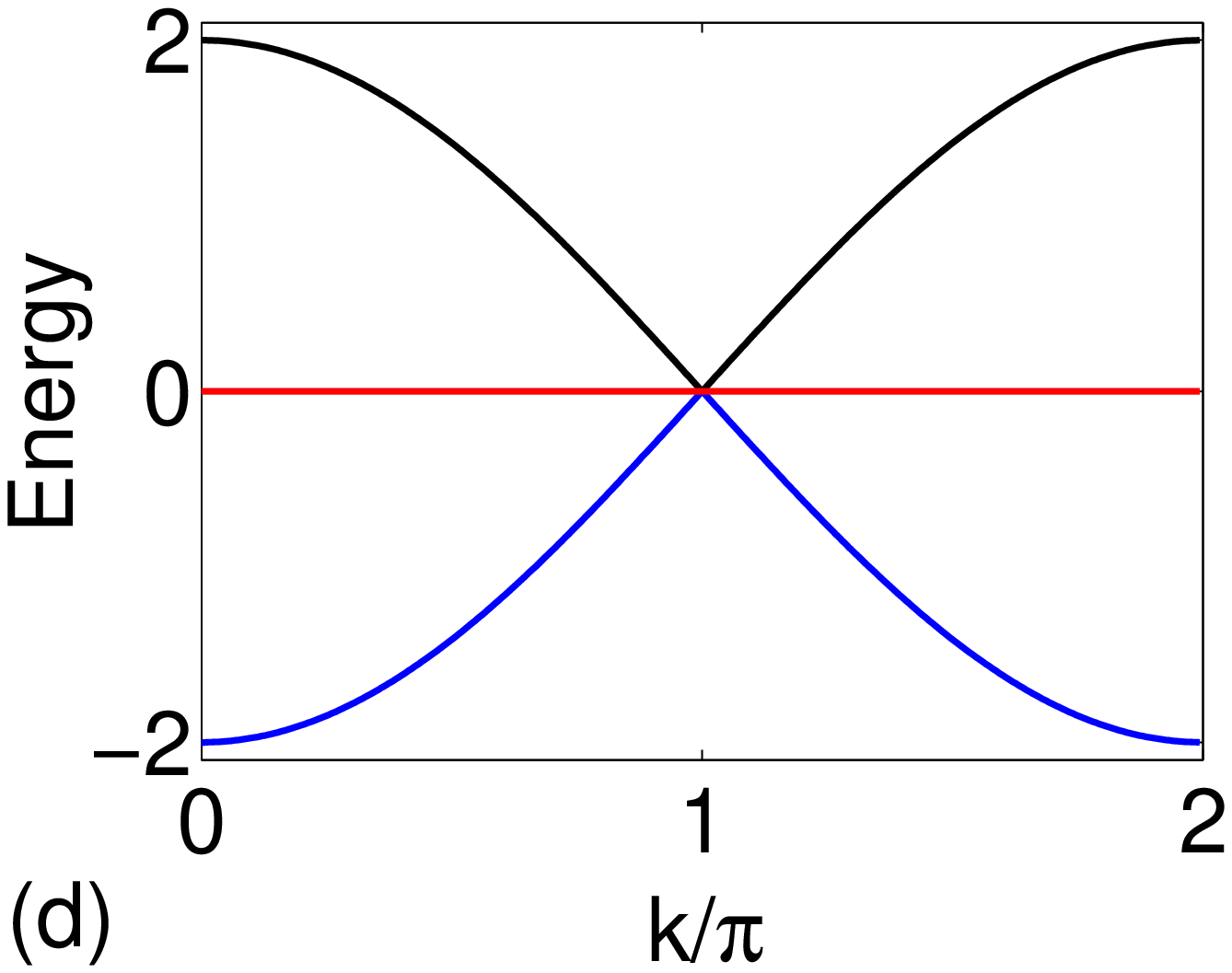}}
\end{minipage}
\caption{(Color online) Energy spectra from Eq. (\protect\ref{spectrum}) at
typical points (a) $g_{e}=0.5$, $g_{o}=0.5$, $\protect\lambda =0.8;$ (b) $%
g_{e}=0.5$, $g_{o}=0.5$, $\protect\lambda =1;$(c) $g_{e}=0$, $g_{o}=0$, $%
\protect\lambda =0.8;$(d) $g_{e}=0$, $g_{o}=0$, $\protect\lambda =1$. We see
that (a) in general, there are four energy bands with symmetry about zero
energy; (b) the two positive (negative) bands touch each other at a single
point when $g_{e}=\pm g_{o}$; (c) when $g_{e}=g_{o} =0,$\ two energy bands
near the zero point become a flat band; (d) the top and the bottom bands
touch each other at a single point at the flat band. }
\label{fig2}
\end{figure}

\section{Band structure and symmetry}

\label{Band structure and symmetry}

Based on the solution of $h_{k}$, all the properties of the system can be
obtained. Now we focus on the property of the eigenstates of $h_{k}$. We
introduce a particle-hole transformation $\mathcal{P}$, which is defined as

\begin{equation}
\mathcal{P}^{-1}c_{l}\mathcal{P}=c_{l}^{\dagger }.
\end{equation}%
Applying $\mathcal{P}$\ on operators $\alpha _{k}$\ and $\beta _{k}$\ we
immediately have
\begin{equation}
\mathcal{P}^{-1}\alpha _{k}\mathcal{P}=\alpha _{-k}^{\dagger },\mathcal{P}%
^{-1}\beta _{k}\mathcal{P}=\beta _{-k}^{\dagger },
\end{equation}%
and%
\begin{equation}
\mathcal{P}^{-1}\psi _{k}\mathcal{P}=C\psi _{k},
\end{equation}%
where%
\begin{equation}
C=\left(
\begin{array}{cccc}
-1 & 0 & 0 & 0 \\
0 & -1 & 0 & 0 \\
0 & 0 & 1 & 0 \\
0 & 0 & 0 & 1%
\end{array}%
\right) .
\end{equation}%
It is easy to check that
\begin{equation}
\{C,h_{k}\}=0,
\end{equation}%
i.e., $h_{k}$\ has the particle-hole symmetry. From the secular equation $%
h_{k}\left\vert \phi _{\rho \sigma }^{k}\right\rangle =\varepsilon _{\rho
\sigma }^{k}\left\vert \phi _{\rho \sigma }^{k}\right\rangle $, we have%
\begin{equation}
C\left\vert \phi _{\rho \sigma }^{k}\right\rangle =\left\vert \phi _{\bar{%
\rho}\sigma }^{k}\right\rangle ,
\end{equation}%
with labels $\overline{\sigma }=-\sigma $ and $\overline{\rho }=-\rho $,
which accords with the expression of $\left\vert \phi _{\rho \sigma
}^{k}\right\rangle $\ in Eq. (\ref{vectors}). Then we have%
\begin{equation}
\left\langle \phi _{\rho \sigma }^{k}\right\vert \partial _{k}\left\vert
\phi _{\rho \sigma }^{k}\right\rangle =\left\langle \phi _{\bar{\rho}\sigma
}^{k}\right\vert \partial _{k}\left\vert \phi _{\bar{\rho}\sigma
}^{k}\right\rangle ,  \label{Berry connection1}
\end{equation}%
which indiates that the particle state $\left\vert \phi _{+\sigma
}^{k}\right\rangle $\ and hole state $\left\vert \phi _{-\sigma
}^{k}\right\rangle $ have the identical Berry connection.

We note that $C$\ and $h_{k}$\ cannot have the common eigenstates. However,
we have
\begin{equation}
\lbrack C,(h_{k})^{2}]=0,
\end{equation}%
which means that $C$ and $(h_{k})^{2}$\ can have the same eigenstates. In
other word, matrix\ $(h_{k})^{2}$\ can be block diagonalized in the $%
\mathcal{P}$ symmetry invariant subspaces. Actually it is easy to check that

\begin{equation}
2(h_{k})^{2}=\left(
\begin{array}{cc}
h_{k}^{e} & 0 \\
0 & h_{k}^{o}%
\end{array}%
\right) +\Lambda _{k},
\end{equation}%
where%
\begin{equation}
h_{k}^{e}=\left(
\begin{array}{cc}
\Lambda _{k}-2g_{e}^{2} & -2\gamma _{-k}g_{e} \\
-2\gamma _{k}g_{e} & 2g_{e}^{2}-\Lambda _{k}%
\end{array}%
\right) ,
\end{equation}%
and%
\begin{equation}
h_{k}^{o}=\left(
\begin{array}{cc}
\Lambda _{k}-2g_{o}^{2} & 2\gamma _{k}g_{o} \\
2\gamma _{-k}g_{o} & 2g_{o}^{2}-\Lambda _{k}%
\end{array}%
\right) .
\end{equation}%
Straightforward derivation shows that matrices $h_{k}^{e}$\ and $h_{k}^{o}$\
have the relations: (i)
\begin{equation}
\sigma _{y}h_{k}^{e}\sigma _{y}=-h_{-k}^{e},\sigma _{y}h_{k}^{o}\sigma
_{y}=-h_{-k}^{o},
\end{equation}%
with%
\begin{equation}
\sigma _{y}=\left(
\begin{array}{cc}
0 & -i \\
i & 0%
\end{array}%
\right) ;
\end{equation}%
(ii) they have the same eigenvalues
\begin{equation}
\epsilon _{\sigma }^{k}=2(\varepsilon _{\rho \sigma }^{k})^{2}-\Lambda _{k}.
\label{reduced spectrum}
\end{equation}%
It indicates that, the four bands can be reduced to two bands. For an
arbitrary state in the form%
\begin{equation}
\left\vert \Phi _{\sigma }^{k}\right\rangle =\alpha \left\vert \phi _{\sigma
\sigma }^{k}\right\rangle +\beta |\phi _{\bar{\sigma}\sigma }^{k}\rangle ,
\label{a_b_}
\end{equation}%
we always have%
\begin{equation}
\left[ 2(h_{k})^{2}-(g_{e}^{2}+g_{o}^{2}+|\gamma _{k}|^{2})\right]
\left\vert \Phi _{\sigma }^{k}\right\rangle =\epsilon _{\sigma
}^{k}\left\vert \Phi _{\sigma }^{k}\right\rangle .  \label{secular eq}
\end{equation}%
Furthermore, applying operator%
\begin{equation}
\Sigma =\left(
\begin{array}{cc}
\sigma _{y} & 0 \\
0 & -\sigma _{y}%
\end{array}%
\right) ,
\end{equation}%
on the Eq. (\ref{secular eq}), we have%
\begin{eqnarray}
\epsilon _{\bar{\sigma}}^{k}\Sigma \left\vert \Phi _{\sigma
}^{k}\right\rangle &=&-\epsilon _{\sigma }^{k}\Sigma \left\vert \Phi
_{\sigma }^{k}\right\rangle  \notag \\
&=&\left[ 2(h_{-k})^{2}-(g_{e}^{2}+g_{o}^{2}+|\gamma _{k}|^{2})\right]
\Sigma \left\vert \Phi _{\sigma }^{k}\right\rangle .  \label{SIGMA}
\end{eqnarray}%
Then we have%
\begin{equation}
\Sigma \left\vert \Phi _{\sigma }^{k}\right\rangle =\alpha ^{\prime
}\left\vert \phi _{\bar{\sigma}\bar{\sigma}}^{-k}\right\rangle +\beta
^{\prime }|\phi _{\sigma \bar{\sigma}}^{-k}\rangle =\left\vert \Phi _{\bar{%
\sigma}}^{\prime -k}\right\rangle ,
\end{equation}%
which reveals the relation between $\left\vert \Phi _{\sigma
}^{k}\right\rangle $ and the eigenstate $\left\vert \Phi _{\overline{\sigma }%
}^{\prime -k}\right\rangle $\ of $(h_{-k})^{2}$. Here factors $\alpha
^{\prime }$\ and $\beta ^{\prime }$\ is obtained as $\alpha $ and $\beta $
from the expression of \ in Eq. (\ref{a_b_}).\ We have another relation
about the Berry connection

\begin{equation}
\left\langle \phi _{\rho \rho }^{k}\right\vert \partial _{k}\left\vert \phi
_{\rho \rho }^{k}\right\rangle =\left\langle \phi _{\rho \bar{\rho}%
}^{-k}\right\vert \partial _{k}\left\vert \phi _{\rho \bar{\rho}%
}^{-k}\right\rangle .  \label{Berry connection2}
\end{equation}%
These results are the basement for exploring the topological feature of the
system.

In addition, two $2\times 2$\ matrices $h_{k}^{e}$\ and $h_{k}^{o}$\
characterize the same physics of the $4\times 4$\ matrix $h_{k}$\ with a
simpler form. For example, in the planes of $g_{o}=\pm g_{e}$,\textbf{\ }Eq.
(\ref{reduced spectrum}) is reduced to $\left\vert \lambda
+e^{ik}\right\vert \sqrt{\left\vert \lambda +e^{ik}\right\vert
^{2}+4g_{o}^{2}}$.\ According to the method developed in Ref. \cite{ZG
PRL,ZG1,LC1}, four nodal lines, are topological quantum-phase boundaries.
Actually, such a dispersion relation links to a loop with parameter equation%
\begin{equation}
\left\{
\begin{array}{c}
x=B_{0}(\lambda +\cos k) \\
y=B_{0}\sin k%
\end{array}%
\right. ,
\end{equation}%
where $B_{0}=\sqrt{\lambda ^{2}+2\lambda \cos k+1+4g_{o}^{2}}$. The winding
number of the loop%
\begin{equation}
v=\left\{
\begin{array}{cc}
1, & \left\vert \lambda \right\vert <1 \\
0, & \left\vert \lambda \right\vert >1%
\end{array}%
\right. ,
\end{equation}%
which identify the topological invariants of the quantum phases.

\section{Topological invariants}

\label{Topological invariants}

The aim of this paper is to investigate the topological feature of the nodal
lines in the three dimensional parameter\ space. The topological properties
of $1$D system are characterized by the so-called Zak phase, the Berry's
phase picked up by an eigenstate of $h_{k}$\ sweeping across the Brillouin
zone. There are four eigenstates for a given $k$. The Zak phase for the
branch of $\left\vert \phi _{\rho \sigma }^{k}\right\rangle $\ is%
\begin{equation}
\mathcal{Z}_{\rho \sigma }(g_{e},g_{o},\lambda )=\frac{1}{2\pi }\int_{-\pi
}^{\pi }\left\langle \phi _{\rho \sigma }^{k}\right\vert \frac{\partial }{%
\partial k}\left\vert \phi _{\rho \sigma }^{k}\right\rangle dk.
\end{equation}%
The equations of the Berry connection (\ref{Berry connection1}) and (\ref%
{Berry connection2}) show that%
\begin{equation}
\mathcal{Z}_{\rho \sigma }=\mathcal{Z}_{\bar{\rho}\sigma },\mathcal{Z}_{\rho
\rho }=-\mathcal{Z}_{\rho \bar{\rho}}
\end{equation}%
which allow us only consider the property of the Zak phase $\mathcal{Z}_{++}$%
. In the following we neglect the subscript by taking $\mathcal{Z}_{++}=%
\mathcal{Z}$.

It has been shown that quantity

\begin{equation}
\mathcal{N}=\oint_{L}\mathbf{\nabla }\mathcal{Z}\cdot d\mathbf{r},
\end{equation}%
is equivalent to Chern number, which is quantized as a topological
invariant, characterizing the degenerate point in the parameter space. Here $%
\mathbf{\nabla }$\ is the nabla operator%
\begin{equation}
\mathbf{\nabla }=(\frac{\partial }{\partial g_{e}}\mathbf{i}+\frac{\partial
}{\partial g_{o}}\mathbf{j}+\frac{\partial }{\partial \lambda }\mathbf{k}),
\end{equation}%
and $r=g_{e}i+g_{o}j+\lambda k$\ is the vector in the parameter space. The
value of $N$\ depends on the topology of loop $L$\ in the parameter space: $%
N $\ is zero if the loop does not encircle the nodal line, while may be
nonzero if encircle the nodal line.\ However, it is a little complicated to
determine the exact integer $N$\ from the direct derivation from state $%
\left\vert \phi _{\rho \sigma }^{k}\right\rangle $. Nevertheless, all the
loops have the same $N$, if they encircle the identical nodal lines.
Therefore, each of semi-infinite lines has their own field current,
indicating the Chern number encircling them. In order to determine the Chern
number, we rewrite $h_{k}$ as the form

\begin{equation}
h_{k}=\left(
\begin{array}{cccc}
2g_{o} & \gamma _{-k} & 0 & \gamma _{-k} \\
\gamma _{k} & 2g_{_{e}} & -\gamma _{k} & 0 \\
0 & -\gamma _{-k} & -2g_{o} & -\gamma _{-k} \\
\gamma _{k} & 0 & -\gamma _{k} & -2g_{_{e}}%
\end{array}%
\right) ,
\end{equation}%
in the operator basis
\begin{equation}
\psi _{k}^{\dag }=\frac{1}{\sqrt{2}}\left(
\begin{array}{cccc}
-\beta _{-k}, & \alpha _{-k}, & -\beta _{k}^{\dagger }, & \alpha _{k}^{\dag }%
\end{array}%
\right) .
\end{equation}%
In the case of $g_{o}\approx g_{e}\gg 1\gg \left\vert 1-\lambda \right\vert $%
, we have

\begin{equation}
h_{k}\approx \left(
\begin{array}{cc}
h_{k}^{\mathrm{RM}} & 0 \\
0 & -h_{k}^{\mathrm{RM}}%
\end{array}%
\right) +\left( g_{o}+g_{e}\right) \left(
\begin{array}{cc}
I_{2} & 0 \\
0 & -I_{2}%
\end{array}%
\right) ,
\end{equation}%
where the core matrix%
\begin{equation}
h_{k}^{\mathrm{RM}}=\left(
\begin{array}{cc}
\Delta & \gamma _{-k} \\
\gamma _{k} & -\Delta%
\end{array}%
\right) ,
\end{equation}%
is nothing but the Bloch Hamiltonian of a standard RM model \cite{Rice},
with a staggered on-site potential $\Delta =g_{o}-g_{e}$.

It has been shown that if the system adiabatically evolves along a loop
enclosing the degeneracy points $(0,\pm 1)$ in the $\Delta -\lambda $ plane,
then the polarization will be changed by $\pm 1$, where the sign depends on
the direction of the loop. On the other hand, if the loop does not contain
the degeneracy point, then the pumped charge is zero \cite{XD}. Similarly,
the direction of each nodal lines can be determined. We demonstrate
topological properties of these lines in Fig. \ref{fig1}, which indicates
that the nodal lines act as current carrying wires, obeying the topology of
the electric circuits.

\begin{figure}[tbp]
\includegraphics[ bb=22 399 497 753, width=0.45\textwidth, clip]{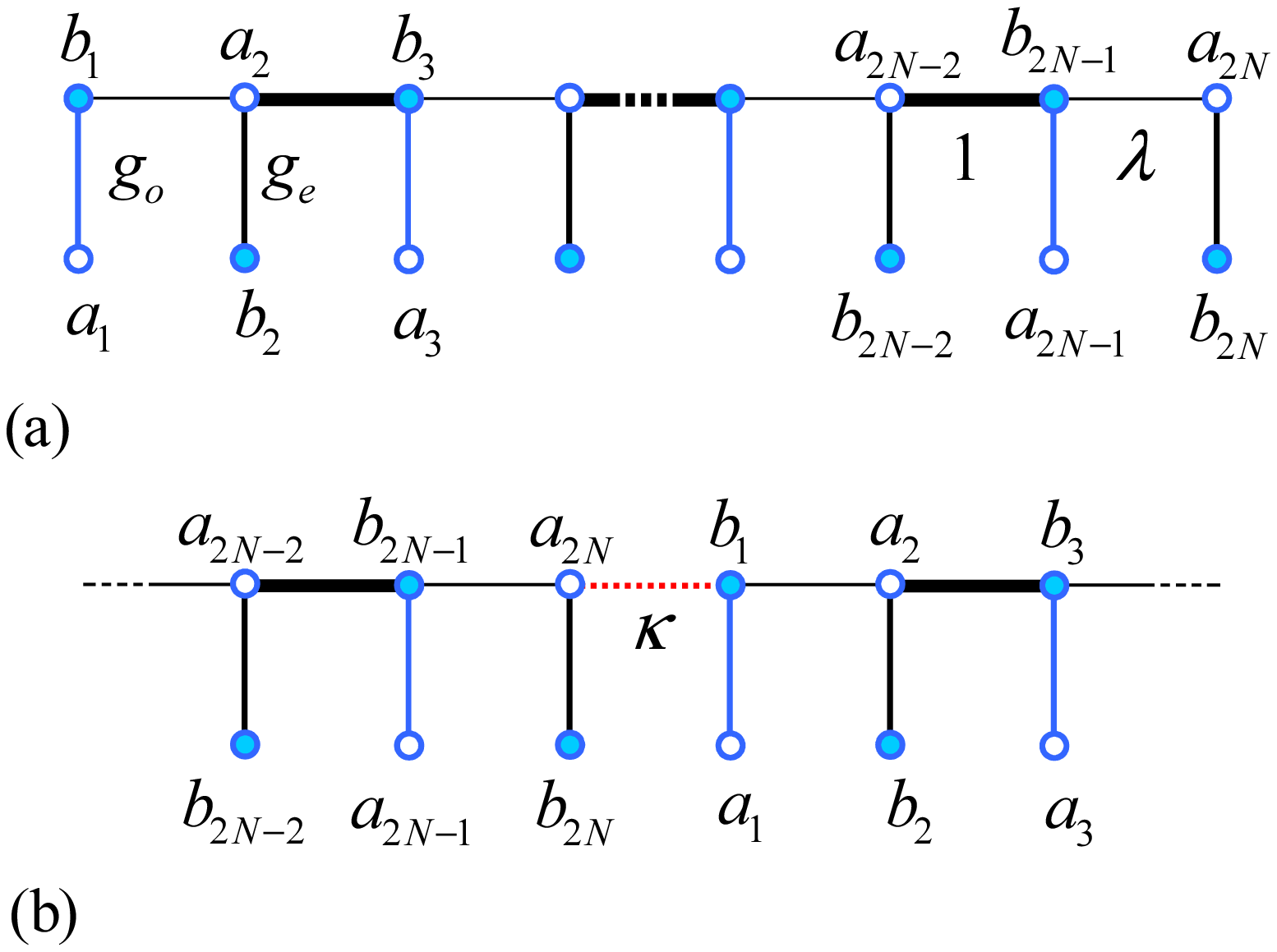}
\caption{(Color online) Lattice geometries for the Majorana models described
in Eqs. (\protect\ref{open}) and (\protect\ref{tunnel}), with open boundary
(a) and weak tunneling $\protect\kappa $ between two ends (b), respectively.
Solid (empty) circle indicates (anti) Majorana modes, while solid and dashed
lines indicate intra-chain and across end to end weak hopping terms,
respectively. Both panels (a) and (b) represent a SSH chain with staggered
side-couplings, forming a comb-like lattice.}
\label{fig3}
\end{figure}

\section{Edge modes and adiabatic transport}

\label{Edge mode and adiabatic transport}
\begin{figure}[tbp]
\includegraphics[ bb=55 196 508 631, width=0.4\textwidth, clip]{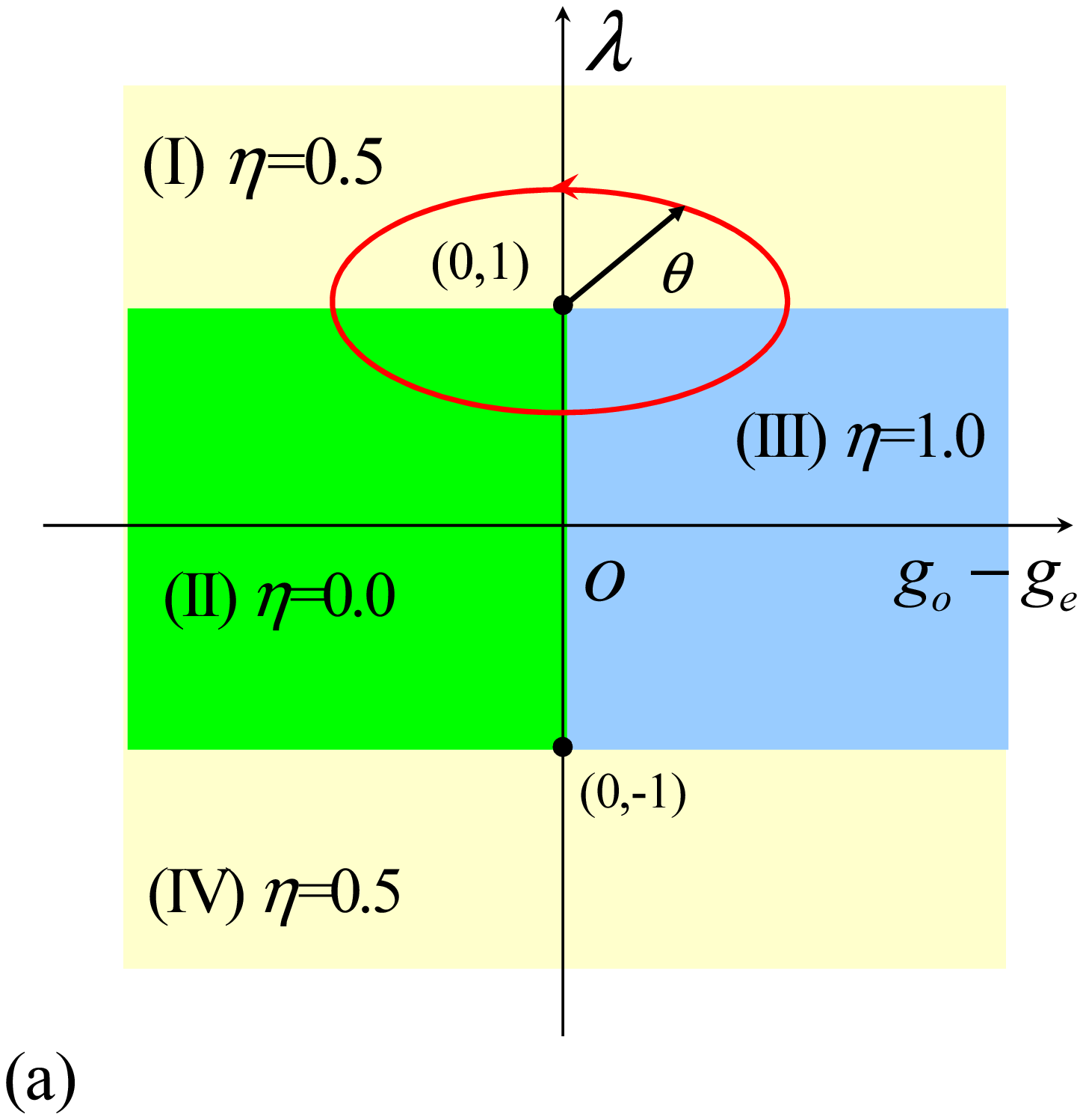} %
\includegraphics[ bb=28 208 540 580, width=0.38\textwidth, clip]{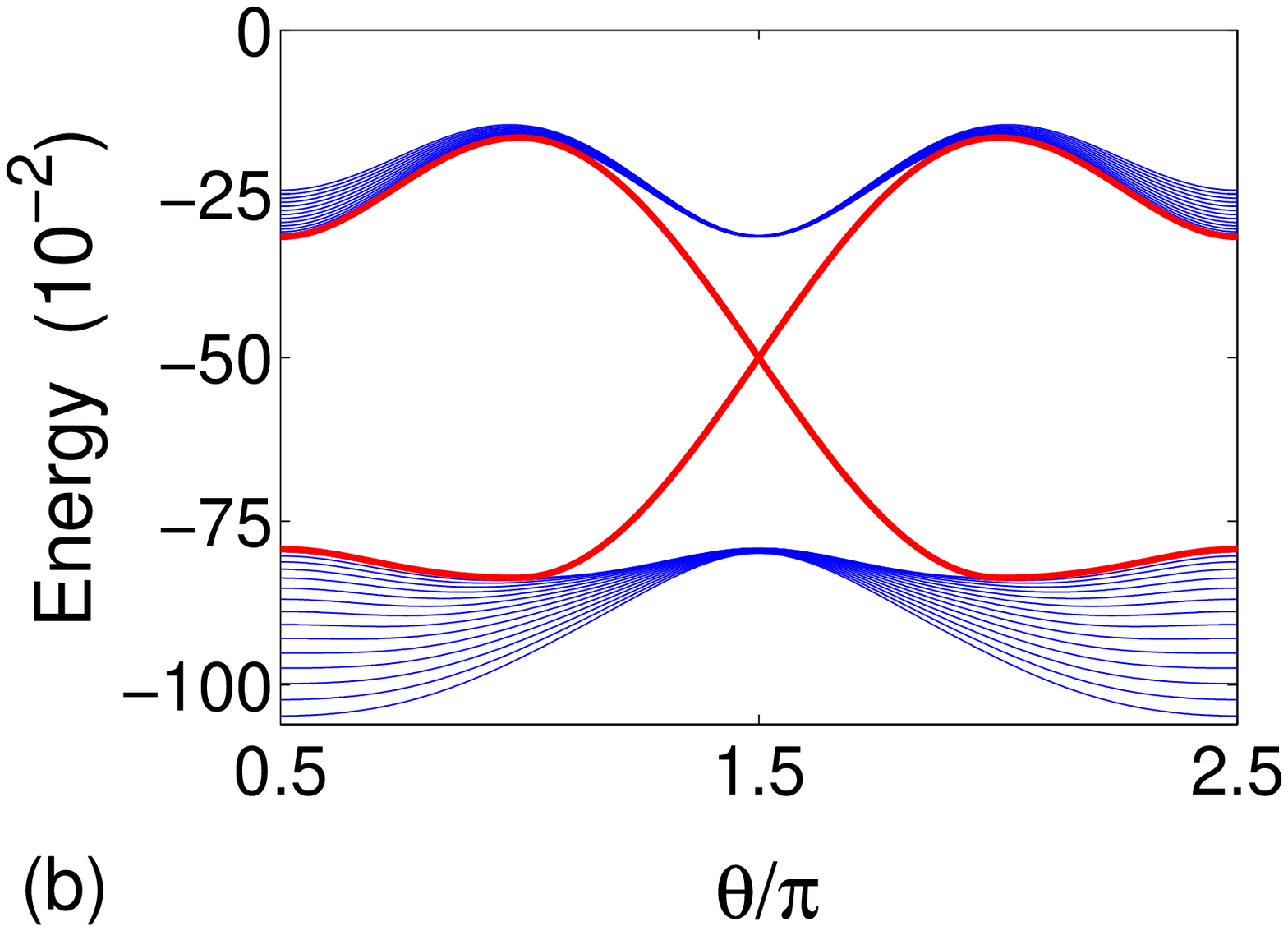} %
\includegraphics[ bb=35 194 548 605, width=0.38\textwidth, clip]{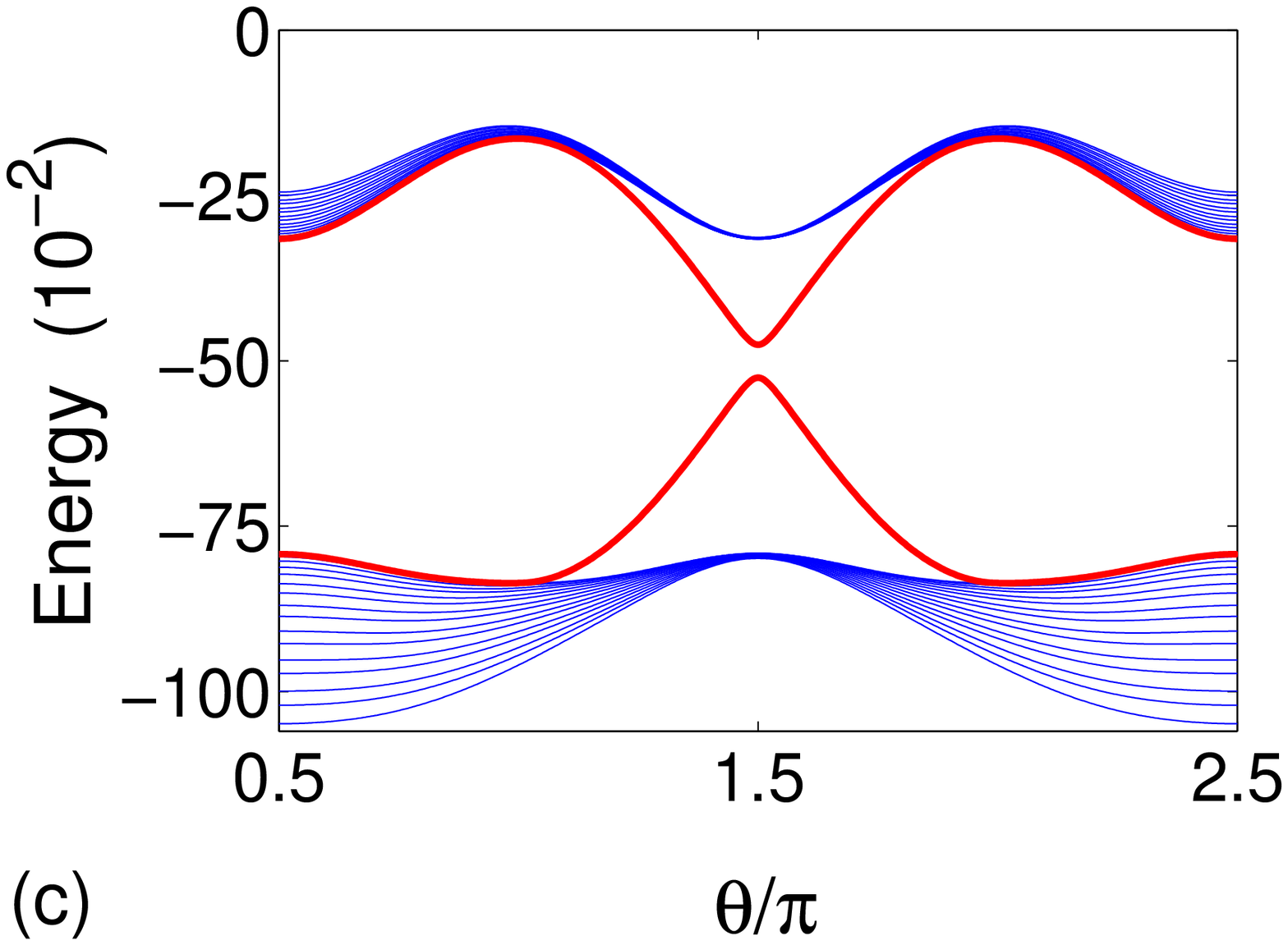}
\caption{(Color online) (a) Majorana dimensionless center of mass
distribution on the parameter space, $(g_{o}-g_{e})-\protect\lambda $ plane
for the mid-gap state in Eq. (\protect\ref{COM}). There are four regions
filled by different colors. If we calculate the change of $\protect\eta $
along a loop enclosing the degeneracy point $(0,\pm 1)$ in the space, then $%
\protect\eta $ will be changed by $\pm 1$. On the other hand, if the loop
does not contain the degeneracy point, then the change of $\protect\eta $ is
zero. (b) and (c) are plots of the spectra of Majorana lattice with
parameters along the cycle in (a) with open and weak link boundary condition
respectively. The spectrum (c) indicates that the change of $\protect\eta $
along a loop can be obtained by an adiabatic process for the mid-gap state
marked in red.}
\label{fig4}
\end{figure}

The above results indicate that the quantum phase boundaries of the model $H$
exhibit the topological characterization. Inspired by the principle of
bulk-edge correspondence\ we expect that the hidden topology behind the
model can be unveiled by exploring the edge modes of the corresponding
Majorana Hamiltonian. In general, the edge modes always\ live at zero energy
\cite{J.A}.\ However, the key feature as a topological invariant is the
bound state, rather than the zero eigenvalue, which may be shifted by a
constant for some specific model. We will see this point from the following
example. Considering the spinless fermion system with an open boundary
condition, the Hamiltonians read%
\begin{eqnarray}
&&H_{\mathrm{CH}}=H-M,  \notag \\
&&M=\left( c_{2N}^{\dagger }c_{1}^{\dagger }+c_{2N}^{\dagger }c_{1}\right) +%
\mathrm{H.c.},
\end{eqnarray}%
which represent the original system with breaking the couplings across
neighboring sites $(2N,1)$.

We introduce Majorana fermion operators%
\begin{equation}
a_{j}=c_{j}^{\dagger }+c_{j},b_{j}=-i\left( c_{j}^{\dagger }-c_{j}\right) ,
\label{ab}
\end{equation}%
which satisfy the relations%
\begin{eqnarray}
\left\{ a_{j},a_{j^{\prime }}\right\} &=&2\delta _{j,j^{\prime }},\left\{
b_{j},b_{j^{\prime }}\right\} =2\delta _{j,j^{\prime }},  \notag \\
\left\{ a_{j},b_{j^{\prime }}\right\} &=&0,a_{j}^{2}=b_{j}^{2}=1.
\end{eqnarray}%
The inverse transformation is
\begin{equation}
c_{j}^{\dagger }=\frac{1}{2}\left( a_{j}+ib_{j}\right) ,c_{j}=\frac{1}{2}%
\left( a_{j}-ib_{j}\right) .
\end{equation}%
Then the Majorana representation of the Hamiltonian is%
\begin{eqnarray}
H_{\mathrm{CH}} &=&\sum\limits_{l=1}^{2N}\frac{i}{2}\left( \mu
_{l}a_{l}b_{l}-\mathrm{H.c.}\right)  \notag \\
&&+\sum\limits_{l=1}^{N-1}\frac{i}{2}\left( \lambda
b_{2l-1}a_{2l}+b_{2l+1}a_{2l}-\mathrm{H.c.}\right)  \notag \\
&&+\frac{i}{2}\left( \lambda b_{2N-1}a_{2N}-\mathrm{H.c.}\right) ,
\end{eqnarray}%
where%
\begin{equation}
\mu _{l}=\left\{
\begin{array}{cc}
g_{e}, & \text{even }l \\
g_{o}, & \text{odd }l%
\end{array}%
\right. .
\end{equation}%
We write down the Hamiltonian in the basis $\varphi ^{T}=(a_{1},$ $b_{1},$ $%
a_{2},$ $b_{2},$ $a_{3},$ $b_{3},$ $...)$ and see that%
\begin{equation}
H_{\mathrm{CH}}=\varphi ^{T}h_{\mathrm{CH}}\varphi ,
\end{equation}%
where $h_{\mathrm{CH}}$\ represents a $4N\times 4N$ matrix. Here matrix $h_{%
\mathrm{CH}}$\ is explicitly written as%
\begin{eqnarray}
h_{\mathrm{CH}} &=&\frac{i}{2}\sum\limits_{l=1}^{2N}\left( \mu
_{l}\left\vert l\right\rangle _{AB}\left\langle l\right\vert -\mathrm{H.c.}%
\right)  \label{open} \\
&&+\frac{i}{2}\sum\limits_{l=1}^{N-1}(\lambda \left\vert 2l-1\right\rangle
_{BA}\left\langle 2l\right\vert +\left\vert 2l+1\right\rangle
_{BA}\left\langle 2l\right\vert  \notag \\
&&-\mathrm{H.c.})+\frac{i}{2}(\lambda \left\vert 2N-1\right\rangle
_{BA}\left\langle 2N\right\vert -\mathrm{H.c.}),  \notag
\end{eqnarray}%
where basis $\left\{ \left\vert l\right\rangle _{A},\left\vert
l\right\rangle _{B},l\in \left[ 1,2N\right] \right\} \ $is an orthonormal
complete set, $_{\alpha }\langle l\left\vert l^{\prime }\right\rangle
_{\beta }=\delta _{ll^{\prime }}\delta _{\alpha \beta }$. The basis array is
$(\left\vert 1\right\rangle _{A},$ $\left\vert 1\right\rangle _{B},$ $%
\left\vert 2\right\rangle _{A},$ $\left\vert 2\right\rangle _{B},$ $%
\left\vert 3\right\rangle _{A},$ $\left\vert 3\right\rangle _{B},$ $...)$,
which accords with $\varphi ^{T}$. Schematic illustrations for structures of
$h_{\mathrm{CH}}$\ are described in Fig. \ref{fig3}. It is expected to have
another quantity which allows us to discern two kinds loops encircling a
nodal line or not (also including different types of nodal lines). To this
end, we introduce the concept of center of mass of Majorana fermion for a
mid-gap state $\left\vert \chi _{\rho \sigma }\right\rangle $, which is
defined as%
\begin{equation}
\eta _{\rho \sigma }=\frac{1}{2N}\sum\limits_{l=1}^{2N}\left\langle \chi
_{\rho \sigma }\right\vert l(\left\vert l\right\rangle _{AA}\left\langle
l\right\vert +\left\vert l\right\rangle _{BB}\left\langle l\right\vert
)\left\vert \chi _{\rho \sigma }\right\rangle .
\end{equation}%
Here we take dimensionless length, and then the range of $\eta _{\rho \sigma
}$\ is $[0,1]$. In the following, we will show that all the mid-gap states
and the corresponding $\eta $\ can be obtained exactly.

For $\left\vert \lambda \right\vert <1$\ the mid-gap states $\left\vert \chi
_{\rho \sigma }\right\rangle $ with eigenvalue $E_{\rho \sigma }$%
\begin{eqnarray}
\left(
\begin{array}{c}
\left\vert \chi _{++}\right\rangle \\
\left\vert \chi _{+-}\right\rangle \\
\left\vert \chi _{-+}\right\rangle \\
\left\vert \chi _{--}\right\rangle%
\end{array}%
\right) &=&\frac{1}{\Omega }\sum\limits_{l=1}^{N}\left(
\begin{array}{c}
\left( -\lambda \right) ^{l-1}(\left\vert 2l-1\right\rangle _{A}-i\left\vert
2l-1\right\rangle _{B}), \\
\left( -\lambda \right) ^{N-l}(\left\vert 2l\right\rangle _{A}-i\left\vert
2l\right\rangle _{B}), \\
\left( -\lambda \right) ^{N-l}(\left\vert 2l\right\rangle _{A}+i\left\vert
2l\right\rangle _{B}), \\
\left( -\lambda \right) ^{l-1}(\left\vert 2l-1\right\rangle _{A}+i\left\vert
2l-1\right\rangle _{B}),%
\end{array}%
\right) ,  \notag \\
2E_{\rho \sigma } &=&\left( g_{o},g_{e},-g_{e},-g_{o}\right) ,
\end{eqnarray}%
where $\Omega =\sqrt{2\left( 1-\lambda ^{2N}\right) /\left( 1-\lambda
^{2}\right) }\approx \sqrt{2/\left( 1-\lambda ^{2}\right) }$. We note that
they are all edged states but not zero mode. The corresponding center of
mass are%
\begin{equation}
\eta =\eta _{\rho \sigma }=\frac{1}{N\Omega ^{2}}\sum\limits_{l=1}^{N}\left%
\{
\begin{array}{cc}
\left( 2l-1\right) \lambda ^{2\left( l-1\right) }, & \rho -\sigma =0 \\
\left( 2l\right) \lambda ^{2\left( N-l\right) }, & \rho +\sigma =0%
\end{array}%
\right. .  \label{COM}
\end{equation}%
It shows that when $\left\vert \lambda \right\vert $\ approaches to $1$, the
mid-gap states close to extended states with a uniform probability
distribution, the center of mass locate at the center of chain. And the
level becomes the band edge in the region $\left\vert \lambda \right\vert
\leqslant 1$. It accords with the plots of band structure of $h_{\mathrm{CH}%
} $\ in Fig. \ref{fig4}.

\begin{figure}[tbp]
\includegraphics[ bb=24 326 539 543, width=0.50\textwidth, clip]{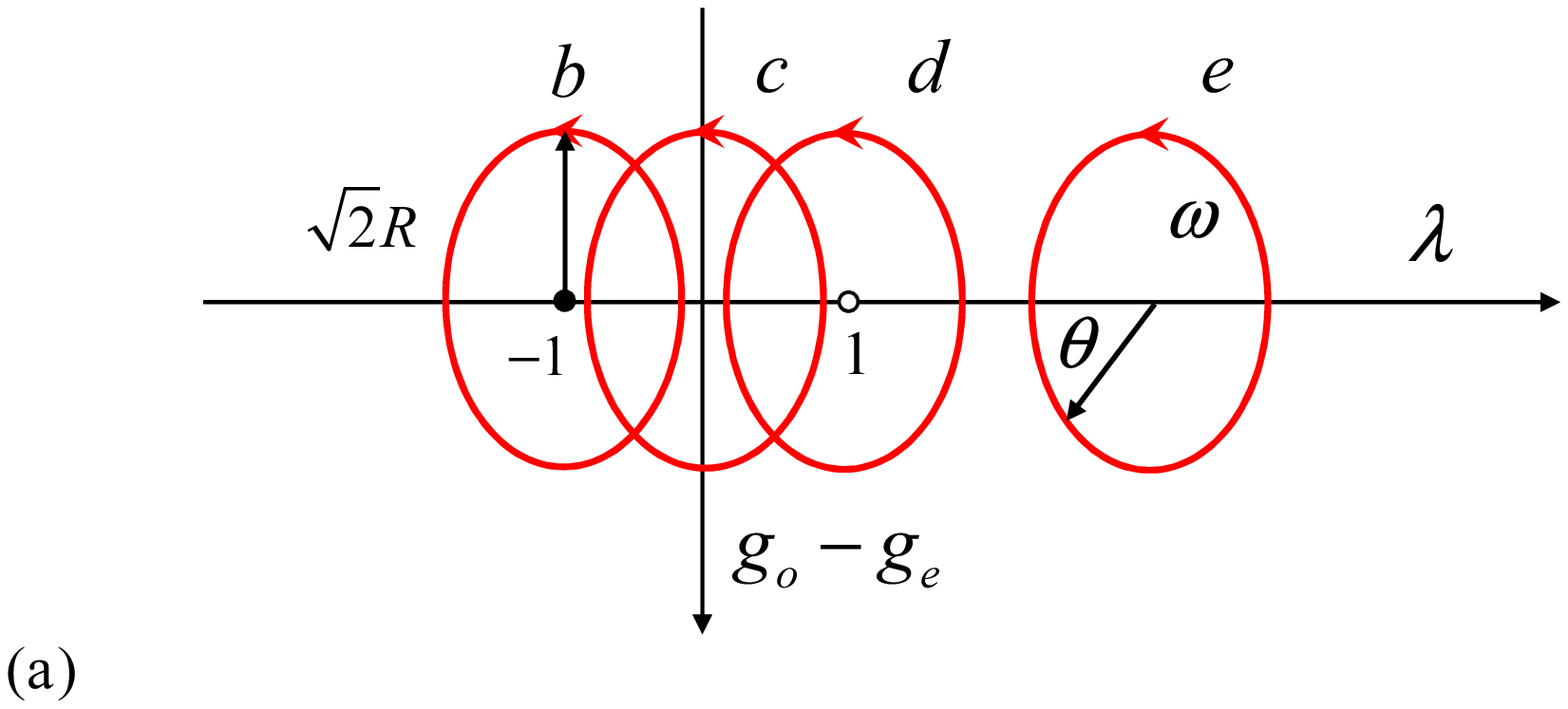}
\begin{minipage}{0.49\linewidth}
\centerline{\includegraphics[width=4.4cm]{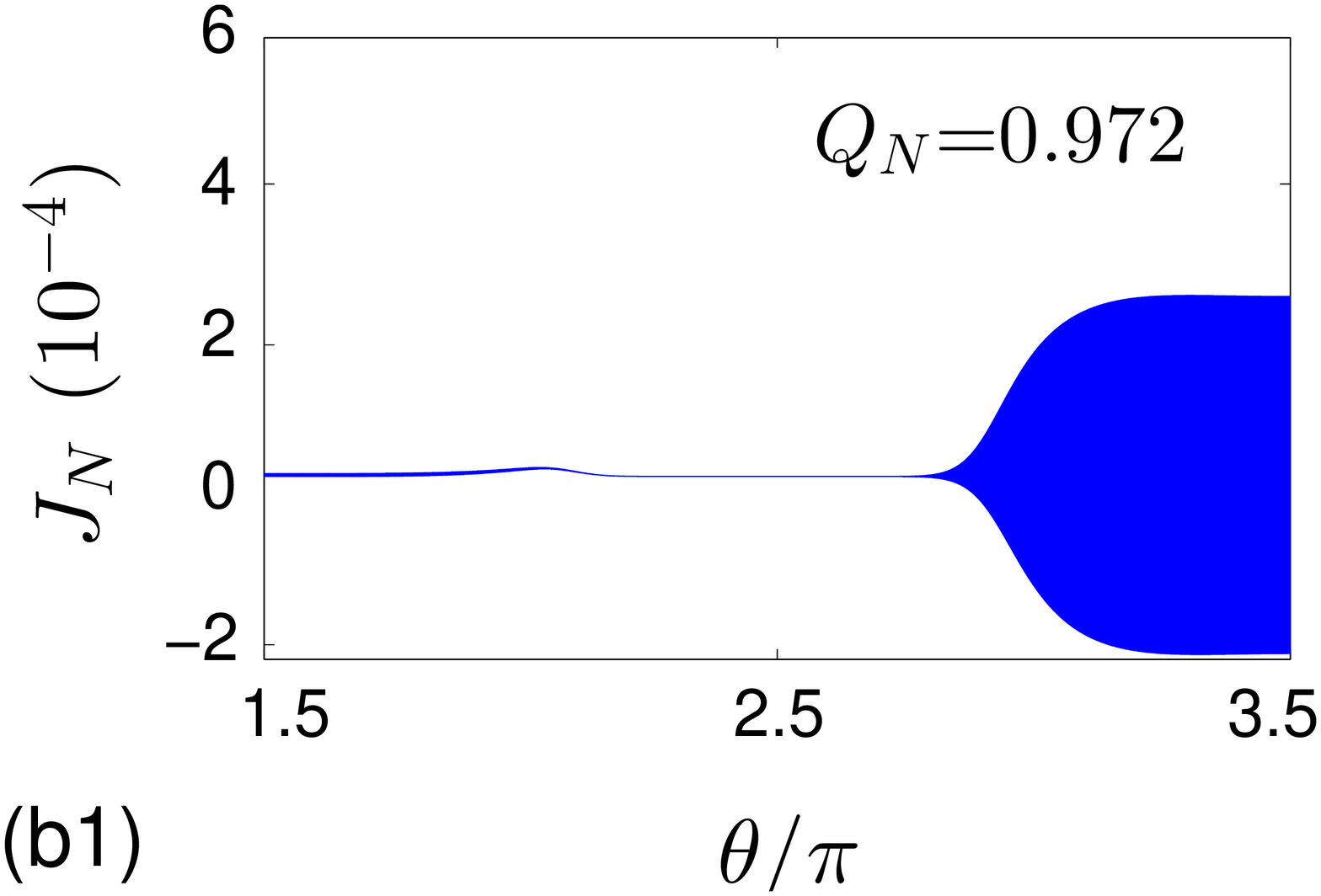}}
\end{minipage}
\begin{minipage}{0.49\linewidth}
\centerline{\includegraphics[width=4.4cm]{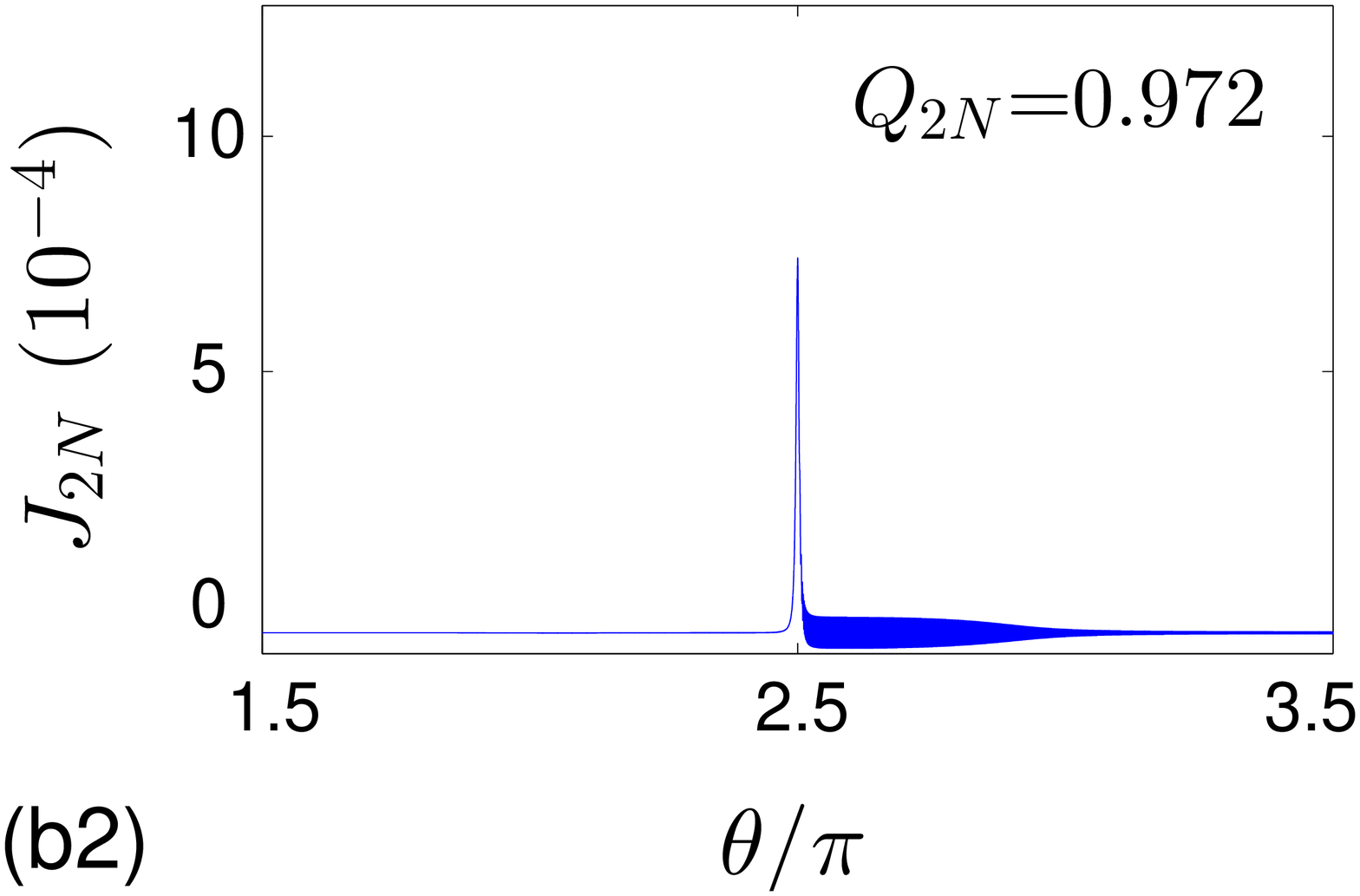}}
\end{minipage}
\begin{minipage}{0.49\linewidth}
\centerline{\includegraphics[width=4.4cm]{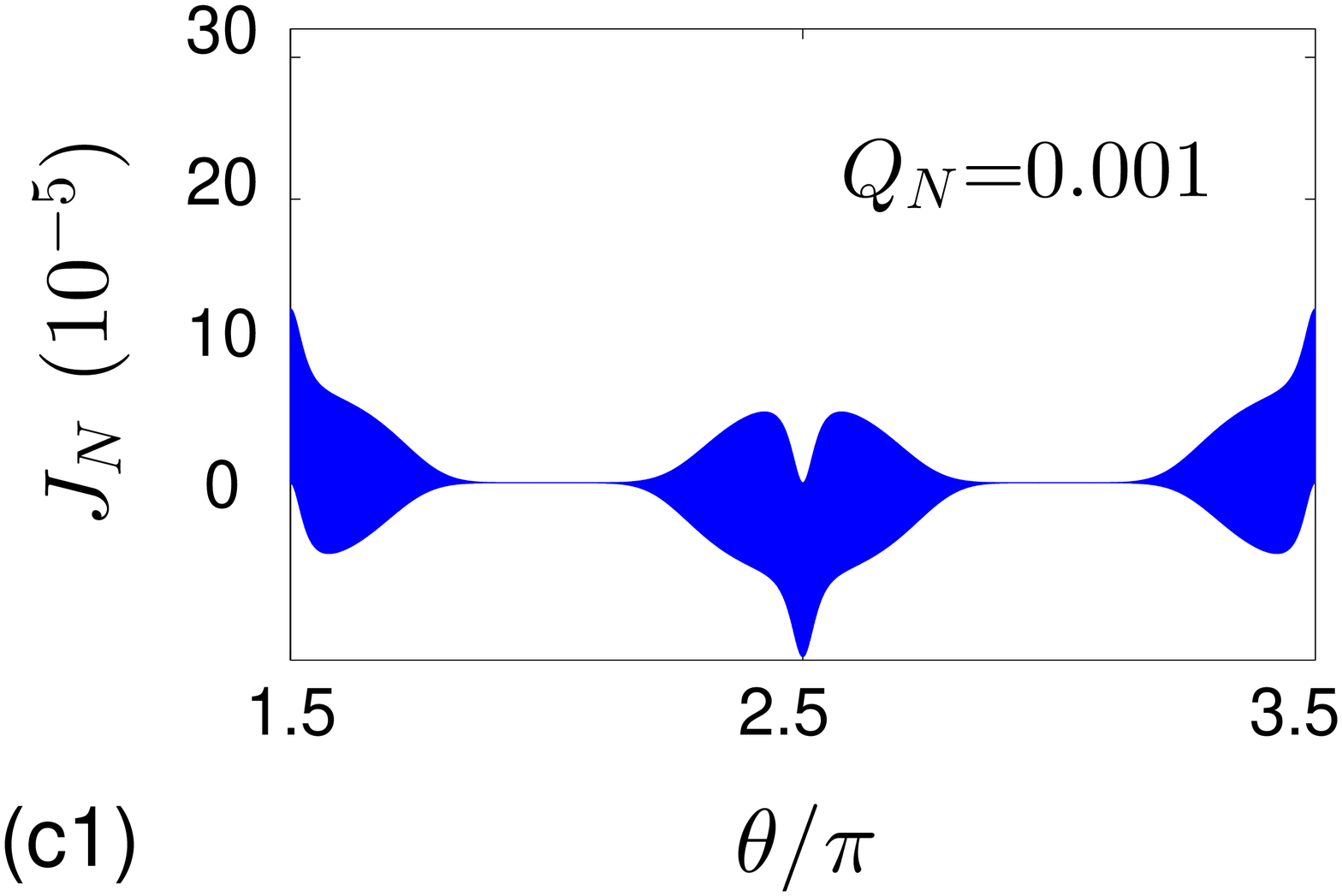}}
\end{minipage}
\begin{minipage}{0.49\linewidth}
\centerline{\includegraphics[width=4.4cm]{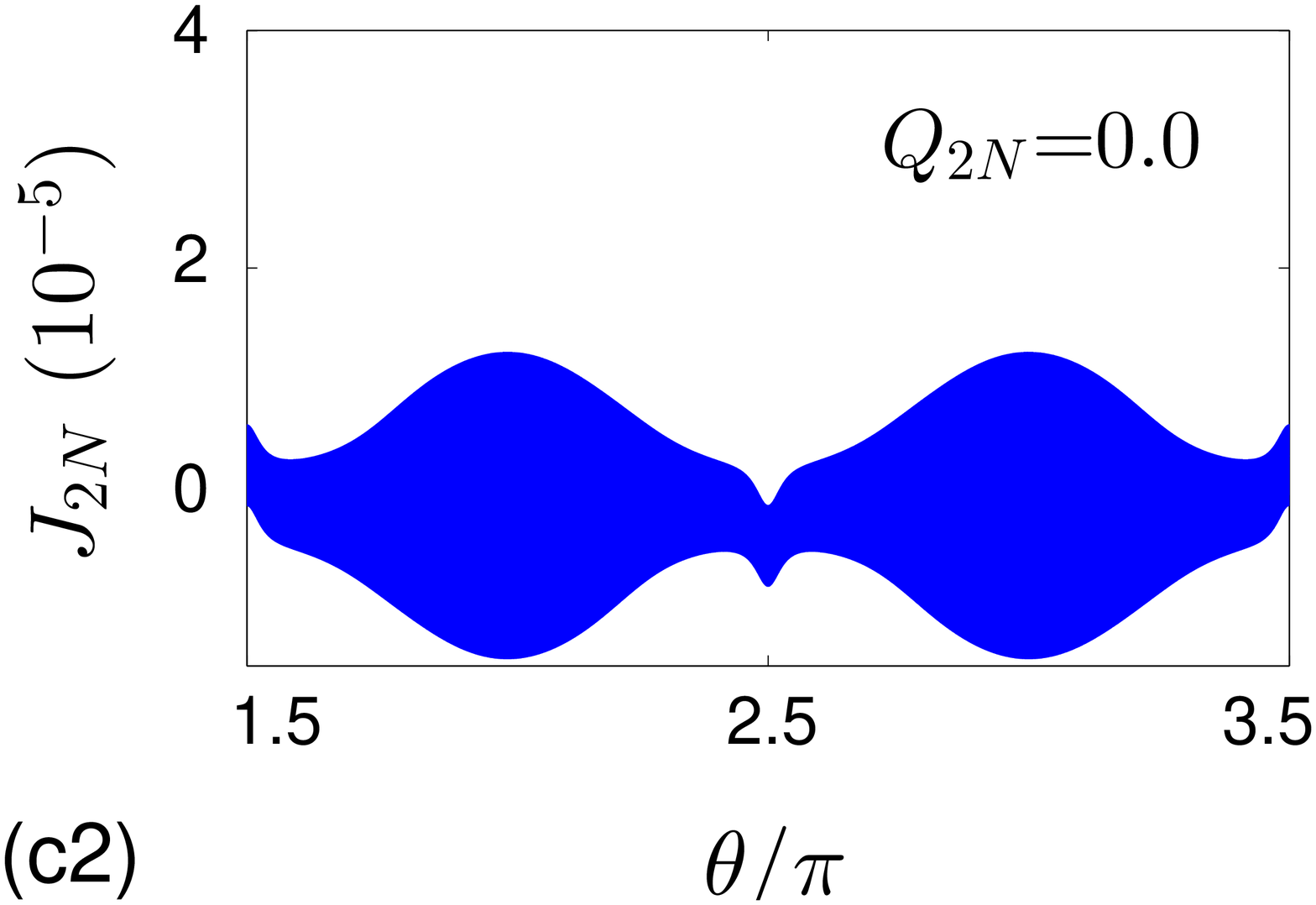}}
\end{minipage}
\begin{minipage}{0.49\linewidth}
\centerline{\includegraphics[width=4.4cm]{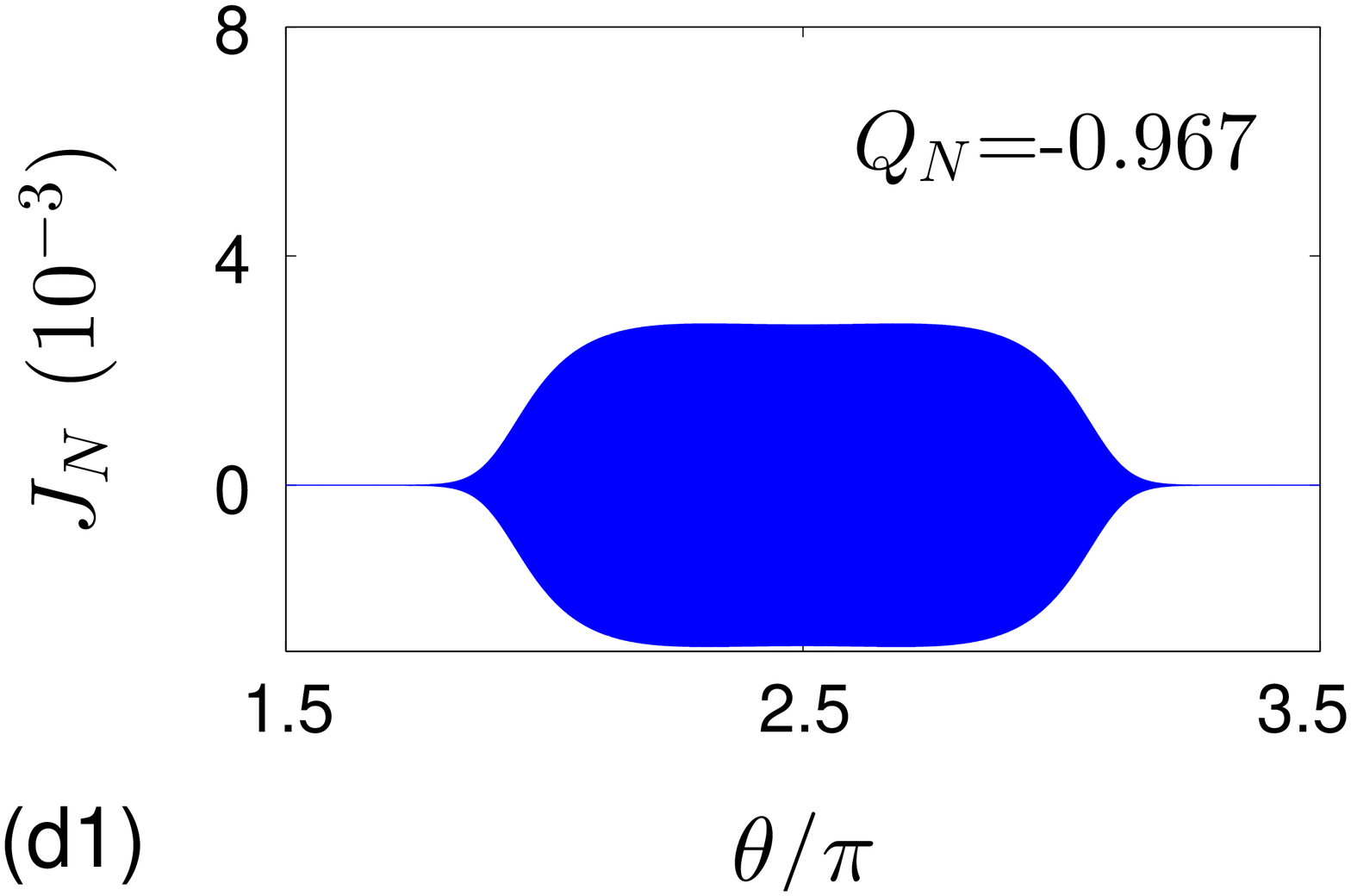}}
\end{minipage}
\begin{minipage}{0.49\linewidth}
\centerline{\includegraphics[width=4.4cm]{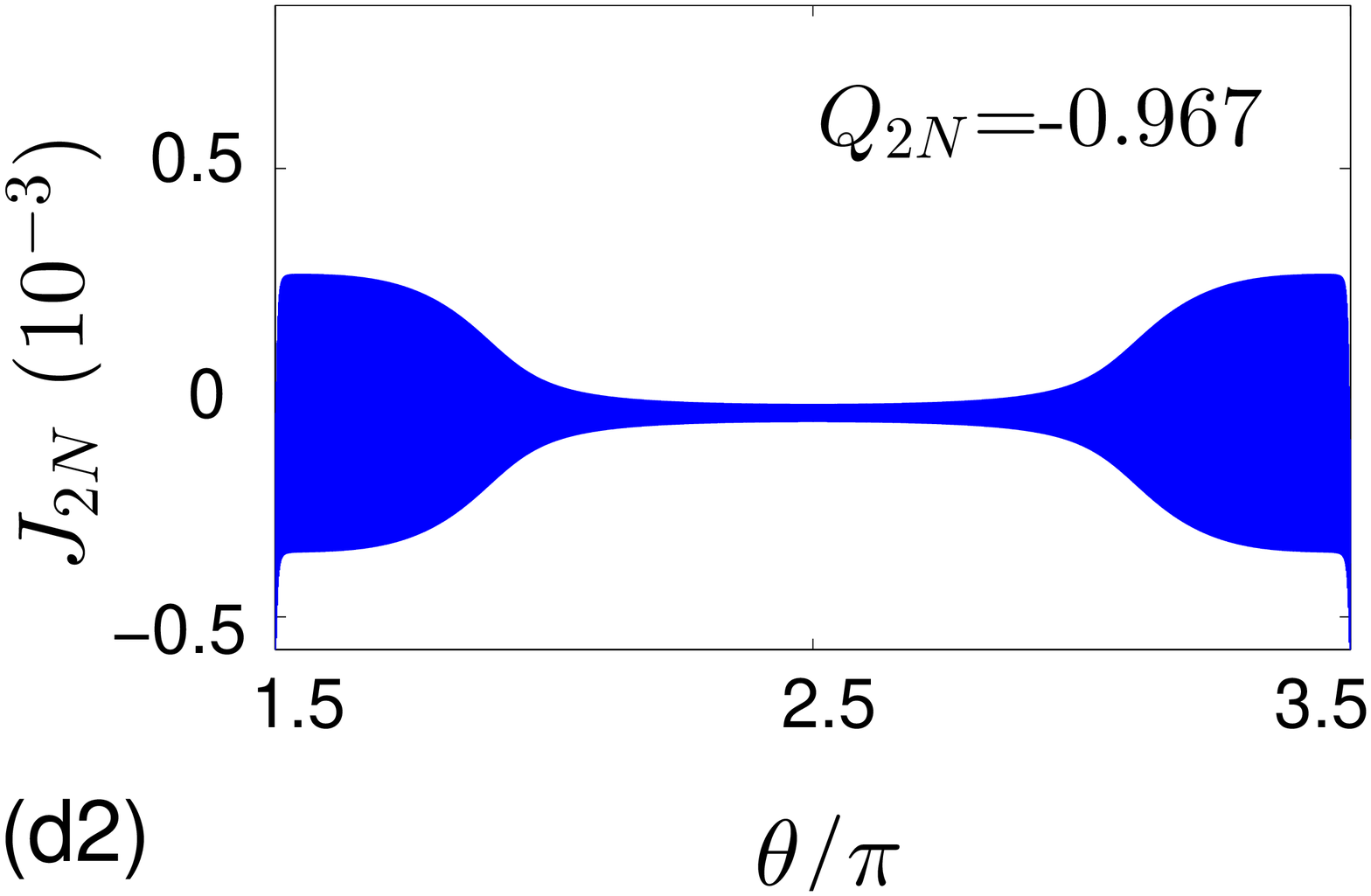}}
\end{minipage}
\begin{minipage}{0.49\linewidth}
\centerline{\includegraphics[width=4.4cm]{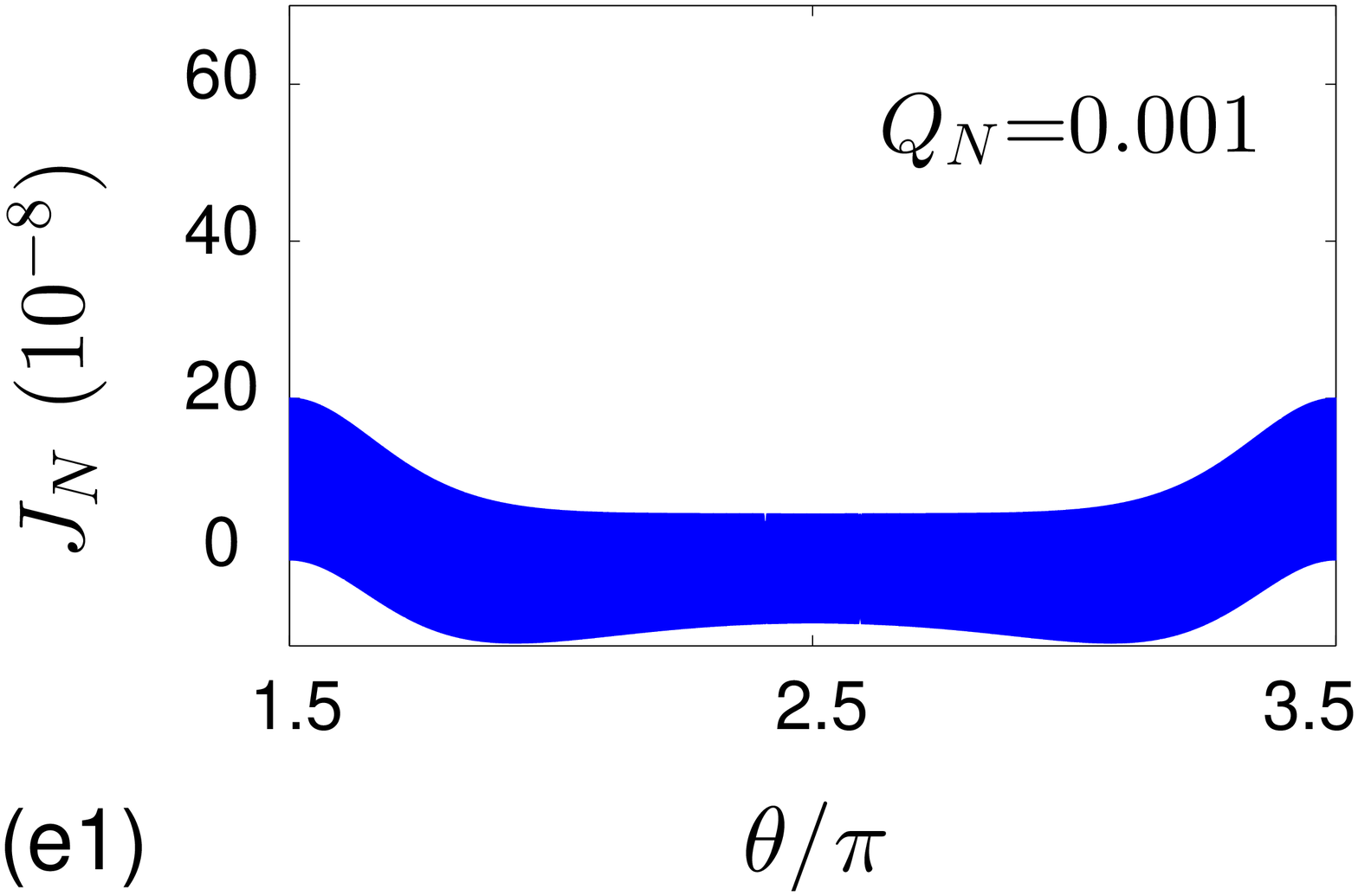}}
\end{minipage}
\begin{minipage}{0.49\linewidth}
\centerline{\includegraphics[width=4.4cm]{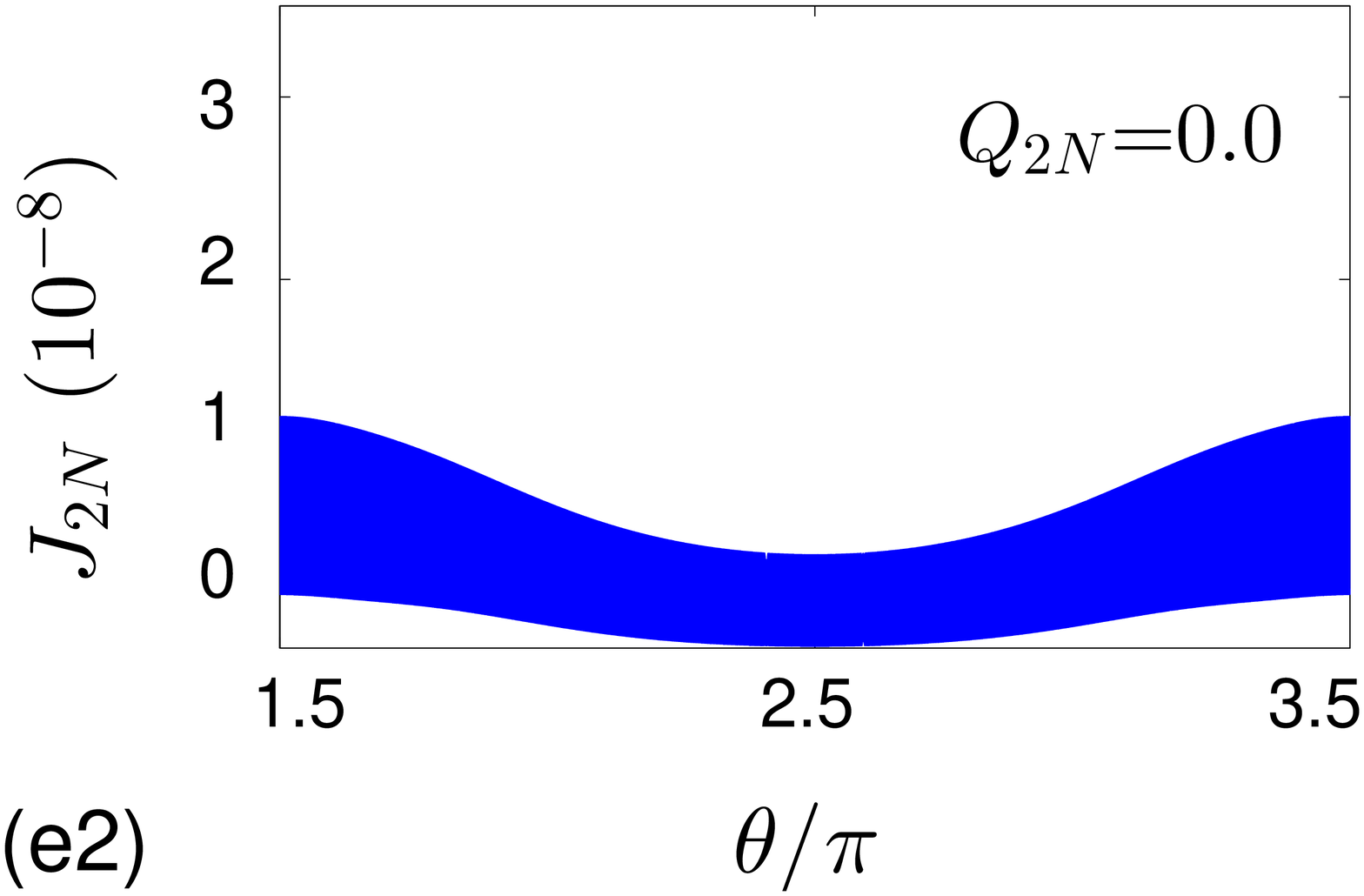}}
\end{minipage}
\caption{(Color online) (a) Schematics of four cycles around several typical
points in the parameter space, $(g_{o}-g_{e})-\protect\lambda $ plane. The
radius $R=0.95$, points $b$ at $(0,-1)$ and $d$ at $(0,1)$ are two
degeneracy points with opposite topological invariants, while points $c $ at
$(0,0)$ and $e$ at $(0,4)$ are two trivial points. (b)-(e) are plots of
Majorana current and the corresponding total probability transfer for
quasi-adiabatic process. The results are obtained numerically for the system
with $N=8$ and $\protect\kappa =0.02$. The speed of time evolution is $%
\protect\omega =5\times 10^{-7}$. It indicates that the topological
invariant can be obtained by dynamical process.}
\label{fig5}
\end{figure}

If we only focus on the mid-gap state just under the first band (shown as
red in Fig. \ref{fig4}(b)). It experiences a sudden change at the vicinity
of $g_{e}=g_{o}$ in the region $\left\vert \lambda \right\vert \leqslant 1$.
Accordingly, there is a jump from $0$ to $1$ for the center of mass. This
behavior is familiar to that of polarization associated with the Zak's phase
\cite{XD}. In general, the topological charge pumping is a transport of
charge through an adiabatic cyclic evolution of the underlying Hamiltonian.
The term topology originates from the fact that the transported charge is
quantized and purely determined by the topology of the pump cycle, making it
robust to perturbation. Furthermore, the pumped charge in each cycle can be
connected to a topological invariant, the Chern number.\ In the present
system, we consider the transport of Majorana fermion. If the system
adiabatically evolves along a loop enclosing the degeneracy line in the $%
(g_{e},g_{o},\lambda )$ space, then the center of mass will be changed by $1$%
, which means that if we allow $(g_{e},g_{o},\lambda )$ to change in time
along this loop, a quantized probability of Majorana fermion is pumped out
of the system after one cycle. It is a similar idea with the Thouless\
topological pump \cite{D.J,Y.H2}. On the other hand, if the loop does not
contains the degeneracy lines, then the pumped probability is zero. The
particle can be pumped out from the left or the right end of the system,
which depends on the direction of the loop. In other word, the pumped
Majorana fermion in each cycle can be connected to its winding number
associated with the field \textbf{P}. To demonstrate this point, we plot the
function $\eta (g_{e},g_{o},\lambda )$\ in the $(g_{o}-g_{e})-\lambda $
plane\ Fig. \ref{fig4}(a). It is clearly that the degeneracy points act as
the vortices of the field $\eta $. \ It indicates that the transport of the
Majorana center of mass directly connect to the topological invariant.

In practice, the pumped probability can be detected by a dynamical process
in a ring system, where the two ends of the Majorana lattice is connected by
a weak coupling. The Hamiltonian has the form

\begin{equation}
h_{\mathrm{R}}=h_{\mathrm{CH}}+\frac{i\kappa _{1,2N}}{2}(\left\vert
1\right\rangle _{BA}\left\langle 2N\right\vert -\mathrm{H.c})  \label{tunnel}
\end{equation}%
where $\kappa $\ is the hopping constant of the weak tunneling. In the
presence of $\kappa $, the spectrum of the Majorana lattice is changed,
which allows the pumped probability flow along the ring. To illustrate this
point, we plot the spectra of $h_{\mathrm{CH}}$\ and $h_{\mathrm{R}}$\
respectively for a circle around a degeneracy point in Fig. \ref{fig4}(b)
and Fig. \ref{fig4}(c).\

When we consider an adiabatic process for a mid-gap state, all the Majorana
probability should pass through two neighboring site $(b_{n},a_{n+1})$ odd
times if the adiabatic passage encloses a degeneracy line, otherwise even
times. The transport of Majorana probability can be witnessed by the current
across $(b_{n},a_{n+1})$, which is defined from the current operator%
\begin{equation}
j_{n}=-\frac{\kappa _{n,n+1}}{2}\left\{
\begin{array}{cc}
(\left\vert n\right\rangle _{BA}\left\langle n+1\right\vert +\mathrm{H.c}),
& n=2l-1 \\
(\left\vert n\right\rangle _{AB}\left\langle n+1\right\vert +\mathrm{H.c}),
& n=2l%
\end{array}%
\right. ,
\end{equation}%
where $l=1,2,...,N$. The total pumped Majorana probability for an
adiabatically evolved mig-gap state through a loop is%
\begin{equation}
Q_{n}=\int_{0}^{T}J_{n}\mathrm{d}t,
\end{equation}%
where $J_{n}=\left\langle \Phi \left( t\right) \right\vert j_{n}\left\vert
\Phi \left( t\right) \right\rangle $, $T$\ is the period of the evolved
time, satisfying $\left\vert \left\langle \Phi \left( t\right) \right\vert
\Phi \left( t+T\right) \rangle \right\vert =1$ for any $t$. The topological
invariant $\mathcal{N}$\ for a given loop corresponds the dynamical quantity
$Q_{n}$, which is independent of $n$. This is referred to dynamical
bulk-edge correspondence, which links the topological feature in bulk to the
dynamical quantity of the corresponding Majorana system.

To examine how the scheme works in practice, we simulate the quasi-adiabatic
process by computing the time evolution numerically for finite system. We
consider the dynamical process along circles in the parameter space, $%
(g_{o}-g_{e})-\lambda $ plane. The path equation is taken as the form%
\begin{equation}
\left\{
\begin{array}{c}
g_{o}-g_{e}=\sqrt{2}R\cos (\omega t) \\
\lambda =\lambda _{\text{c}}+R\sin (\omega t)%
\end{array}%
\right. ,
\end{equation}%
where $(0,\lambda _{\text{c}})$ and $R$ are the center and the semi-minor
axis of the ellipse. For a given initial eigenstate $\left\vert \psi \left(
0\right) \right\rangle $, the time evolved state is%
\begin{equation}
\left\vert \Phi \left( t\right) \right\rangle =\exp (-i\int_{0}^{t}h_{%
\mathrm{R}}\left( t\right) \mathrm{d}t)\left\vert \psi \left( 0\right)
\right\rangle .
\end{equation}%
In low speed limit $\omega \rightarrow 0$, we have%
\begin{equation}
f\left( t\right) =\left\vert \left\langle \Phi \left( t\right) \right\vert
\psi \left( t\right) \rangle \right\vert \rightarrow 1,
\end{equation}%
where $\left\vert \psi \left( t\right) \right\rangle $\ is the corresponding
instantaneous eigenstate of $h_{\mathrm{R}}\left( t\right) $. The
computation is performed by using a uniform mesh in the time discretization
for the time-dependent Hamiltonian $H\left( t\right) $. In order to simulate
a quasi-adiabatic process, we keep $f\left( t\right) >0.997$\ for all the
processes by taking sufficient small $\omega $. Fig. \ref{fig5} plots the
simulations of Majorana current and the corresponding total probability for
several typical cases. It shows that the obtained dynamical quantities are
in closer agreement with the expected values. This paves a way to detect the
topological invariants by dynamical process.

We have established our main results, and a few comments are in order.
First, notice that the zero mode states calculated here are not at exact
zero-energy due to the finite-size of the system ($N=8$). Accordingly, the
Majorana charges are not located at exact edges, i.e., the probability in
the middle of the system is nonzero. However, the numerical results indicate
that the character of topological pump is still evident. This is a benefit
for experimental realization. Second, we would like to point out that the
idea of the topological pump for ultracold atom has been realized
experimentally \cite{S.Na,M.L}.

\section{Summary}

\label{Summary}

In conclusion, we have studied the topological feature of nodal lines in
the\ dimerized Kitaev spin chain with a staggered transverse field. We have
shown that the phase boundary line acts as\ a vortex filament associated
with the field \textbf{P}\ generated by Zak phase. Unlike a topological
quantum phase, which only corresponds to a point in parameter space, a
topological nodal line always involves a loop in parameter space. In
parallel, a nontrivial topological phase always associated with edge states
in the open boundary, while a nontrivial loop is shown to be linked to a
quantized charge pump. By Majorana transformation, we investigated a
Thouless quantum pump with Majorana fermion analytically and numerically.
Our results indicates that the quantized pumping charge can obtained by
quasi-adiabatic circle evolution in a system with an impurity. It provides a
way to observe the dynamical bulk-edge correspondence in experiment.

\acknowledgments We acknowledge the support of the CNSF (Grant No. 11374163).

\end{document}